\documentclass[preprint2]{aastex}
\usepackage{spr-astr-addons}
\usepackage{url}
\usepackage{lscape}
\bibpunct{(}{)}{,}{n}{,}{,}

\RequirePackage{color}

\begin{document}

 \title{Binary stripping as a plausible origin of correlated 
        pairs of extreme trans-Neptunian objects}

 \shorttitle{Binary stripping of ETNOs?}
 \shortauthors{de la Fuente Marcos et al.}

 \author{C.~de~la~Fuente Marcos}
  \and
 \author{R.~de~la~Fuente Marcos}
 \affil{Universidad Complutense de Madrid,
        Ciudad Universitaria, E-28040 Madrid, Spain}
  \and
 \author{S.~J.~Aarseth}
 \affil{Institute of Astronomy, University of Cambridge,
        Madingley Road, Cambridge CB3 0HA, UK
        }
 \email{nbplanet@ucm.es}

  \begin{abstract}
     Asteroids that follow similar orbits may have a dynamical connection as 
     their current paths could be the result of a past interaction with a 
     massive perturber. The pair of extreme trans-Neptunian objects or ETNOs 
     (474640) 2004~VN$_{112}$--2013~RF$_{98}$ exhibits peculiar relative orbital 
     properties, including a difference in longitude of the ascending node of 
     just 1\fdg61 and 3\fdg99 in inclination. In addition, their reflectance 
     spectra are similar in the visible portion of the spectrum. The origin of 
     these similarities remains unclear. Neglecting observational bias, viable 
     scenarios that could explain this level of coincidence include 
     fragmentation and binary dissociation. Here, we present results of 
     extensive direct $N$-body simulations of close encounters between wide 
     binary ETNOs and one trans-Plutonian planet. We find that wide binary 
     ETNOs can dissociate during such interactions and the relative orbital 
     properties of the resulting unbound couples match reasonably well those 
     of several pairs of known ETNOs, including 474640--2013~RF$_{98}$. The 
     possible presence of former binaries among the known ETNOs has strong 
     implications for the interpretation of the observed anisotropies in the 
     distributions of the directions of their orbital poles and perihelia.
  \end{abstract}

  \keywords{Oort Cloud $\cdot$
            Kuiper belt $\cdot$
            Celestial mechanics $\cdot$ 
            Minor planets, asteroids: general $\cdot$
            Minor planets, asteroids: individual: (474640) 2004~VN$_{112}$ $\cdot$
            Minor planets, asteroids: individual: 2013~RF$_{98}$ 
  }

  \section{Introduction}
     The commissioning of large telescopes with wide fields of view equipped with sizeable mosaics of detectors has led to uncovering the 
     presence of a group of extraordinary asteroids whose orbits are larger than those of any other previously known trans-Neptunian object 
     (TNO) and have perihelion distances well outside the range defined by highly-eccentric, already catalogued asteroids and comets. The 
     first member of this fascinating dynamical class was found in 2000, (148209) 2000~CR$_{105}$, and its discovery was soon acknowledged 
     as an important milestone in the study of the outer Solar System because its current path cannot be explained within the standard 
     eight-planets-only Solar System paradigm (e.g. Gladman et al. 2002; Morbidelli and Levison 2004). Sometimes labelled as distant Kuiper 
     belt or inner Oort Cloud objects, Trujillo and Sheppard (2014) called these asteroids extreme TNOs or ETNOs if their semi-major axis, 
     $a$, is greater than 150~AU and their perihelion distance ---$q=a\ (1-e)$, eccentricity, $e$--- is greater than 30~AU. 

     The number of ETNOs included in the database of the Minor Planet Center\footnote{\url{http://www.minorplanetcenter.net/db_search}} 
     (MPC) stands at 28 (as of 2017 August 29), but at least one additional candidate object has already been announced (V774104 by Sheppard 
     et al. 2015). The orbits of the ETNOs do not fit within an eight-planets-only Solar System, but the presence of one or more 
     yet-to-be-discovered planetary bodies orbiting the Sun well beyond Neptune may be able to explain most, if not all, of the unexpected 
     orbital characteristics displayed by the known ETNOs (de la Fuente Marcos and de la Fuente Marcos 2014, 2016a, 2016b, 2016c; Trujillo 
     and Sheppard 2014; de la Fuente Marcos et al. 2015, 2016; Gomes et al. 2015; Batygin and Brown 2016; Brown and Batygin 2016; Malhotra 
     et al. 2016; Sheppard and Trujillo 2016; Millholland and Laughlin 2017). This scenario is often referred to as the trans-Plutonian 
     planets paradigm. As the subject of the possible existence of planets beyond Pluto is far from new, the term trans-Plutonian planet has 
     already been used in the past (see e.g. Seidelmann 1971; Brady 1972; Radzievskij et al. 1994; Kuz'Michev and Tomanov 2006; Lykawka and 
     Mukai 2008); in order to clearly separate TNOs from such planets, hereafter we use trans-Plutonian planet(s) instead of trans-Neptunian 
     planet(s) even if the latter is certainly more usual.

     Perhaps the most popular variant of the trans-Plutonian planets paradigm is the so-called Planet Nine hypothesis (Batygin and Brown 
     2016; Brown and Batygin 2016) that predicts the existence of one $\sim$10~$M_{\oplus}$ planet at about $a=700$~AU based on the analysis 
     of the observed clustering in physical space of the perihelia and the positions of the orbital poles of seven ETNOs ---Sedna, 148209, 
     (474640) 2004~VN$_{112}$, 2007~TG$_{422}$, 2010~GB$_{174}$, 2012~VP$_{113}$ and 2013~RF$_{98}$--- and subsequent analytical and 
     numerical work. Such anisotropies are the by-products of primary clusterings in inclination, $i$, longitude of the ascending node, 
     $\Omega$, and argument of perihelion, $\omega$, as pointed out by de la Fuente Marcos and de la Fuente Marcos (2016c) and, in principle, 
     cannot be attributed to a selection effect (e.g. de la Fuente Marcos and de la Fuente Marcos 2014). However, Shankman et al. (2017a) 
     have claimed that the Planet Nine hypothesis cannot reproduce the overall level of orbital clustering displayed by the known ETNOs and 
     Shankman et al. (2017b) have recently argued that any clustering present among the known ETNOs is probably more apparent than real.  

     The existence of groupings in $\Omega$ and $\omega$ is indicative of some external perturbation only if it can be assumed that there is 
     no detection bias. On the one hand, absence of detection bias is advocated by Trujillo and Sheppard (2014), and by Batygin and Brown 
     (2016), Brown and Batygin (2016) and Brown (2017), while Bannister et al. (2017), Lawler et al. (2017) and Shankman et al. (2017a, 
     2017b) argue that strong detection biases may actually exist. De la Fuente Marcos and de la Fuente Marcos (2014) showed that there is 
     indeed an intrinsic bias in declination, $\delta$, induced by our observing point on Earth; when observed at perihelion or very near it 
     (see their fig. 2), most ETNOs will be discovered at $|\delta|<24${\degr} no matter how complete and extensive the surveys are. This 
     intrinsic detection bias affects the distribution of observed orbital elements (see their fig. 3); in particular, most discoveries 
     should show low orbital inclinations, but this is not what is observed. At this point, the presence or absence of harmful detection 
     biases in the current ETNO sample are both plausible hypotheses until proven otherwise, but investigating this subject is beyond the 
     scope of this paper. 

     Recent observational work has shown that, among the known ETNOs, the pair 474640--2013~RF$_{98}$ stands out in terms of both dynamical 
     and spectroscopic properties, suggesting that these two objects may have had a common physical origin (de Le\'on et al. 2017). The
     hypothesis of the existence of a genetic link for this pair is reasonably well supported by the currently available evidence. If a 
     chance alignment is discarded, viable scenarios that could lead to a pair of closely related minor bodies include fragmentation and 
     binary dissociation. Preliminary calculations show that binary dissociation after an encounter with a trans-Plutonian planet at very 
     large heliocentric distance might be able to explain the origin of this pair of ETNOs (de Le\'on et al. 2017). Here, we present the 
     results of a large number of direct $N$-body simulations of close encounters between wide binary ETNOs and one trans-Plutonian planet 
     aimed at providing a detailed account of the binary dissociation process and its outcome. The goal of this research is not estimating 
     the odds of a binary stripping event happening at hundreds of astronomical units from the Sun for which we do not have enough data yet, 
     but its plausibility; therefore, our numerical exploration is mostly a theoretical exercise with potentially interesting practical 
     applications. This paper is organized as follows. Some relevant properties of the ETNO pair 474640--2013~RF$_{98}$ are summarized in 
     Sect. 2 that also delves into the context of this research. Our $N$-body methodology is briefly outlined in Sect. 3. Section 4 
     describes an extensive exploration of the disruption mechanism and the orbital properties of the resulting unbound couples. The 
     transition from newly disrupted couple to ETNO pair is studied in Sect. 5. Results are discussed in Sect. 6 and conclusions summarized 
     in Sect.~7. 

  \section{The pair (474640)~2004~VN$_\mathbf{112}$--2013~RF$_\mathbf{98}$: relevant data and context}
     The state of the art for the pair (474640) 2004~VN$_{112}$--2013~RF$_{98}$ has been reviewed by de Le\'on et al. (2017). Within the 
     standard eight-planets-only Solar System paradigm and after performing extensive numerical simulations, Sheppard and Trujillo (2016) 
     have classified 474640 as a long-term stable, extreme detached object; this conclusion is consistent with results from an independent 
     analysis carried out by Brown and Batygin (2016). As for the other member of the pair, Sheppard and Trujillo (2016) have classified 
     2013~RF$_{98}$ as an extreme scattered object after finding that it becomes dynamically unstable within the standard paradigm over 10 
     Myr time-scales as a result of the influence of Neptune. In contrast, within the Planet Nine hypothesis (Batygin and Brown 2016; Brown 
     and Batygin 2016) both objects are assumed to be long-term stable although some incarnations of the Planet Nine hypothesis make this 
     pair very unstable on time-scales of order of dozens of Myr, being eventually ejected from the Solar System (see fig. 2 in de la 
     Fuente Marcos et al. 2016). The resonant secular dynamics beyond Neptune in the absence of any significant external perturbers has been 
     systematically explored by Saillenfest et al. (2017); their analysis suggests that the criteria used to consider TNOs as detached from 
     the planets must be revised.

     Prior to 2016 September, the available orbital solutions for this pair of ETNOs gave an angular separation between the directions of 
     their perihelia (those of the vector going from the Sun to the respective perihelion point) of 9\fdg8, very similar to the one between 
     the directions of their velocities at perihelion/aphelion (9\fdg5); however, their orbital poles were much closer at 4\fdg1 and they 
     had similar aphelion distances (589~AU versus 577~AU) as well. Following \"Opik (1971), minor bodies with both similar directions of 
     the orbital poles and perihelia could be part of a group of common physical origin. In an attempt to explore this scenario, astrometry, 
     photometry and visible spectroscopy of the two targets were obtained on 2016 September using the OSIRIS camera-spectrograph at the 
     10.4~m Gran Telescopio Canarias (GTC) telescope (de Le\'on et al. 2017). 

     \subsection{Updated data and their impact}
        The results in de Le\'on et al. (2017) show that the spectral slopes of 474640--2013~RF$_{98}$ are very close matches, similar to 
        those of (148209) 2000~CR$_{105}$ and 2012~VP$_{113}$, and compatible with the ones of 2002~GB$_{32}$ and 2003~HB$_{57}$ (two other 
        members of the ETNO category not linked to the Planet Nine hypothesis). However, they are very different from that of Sedna, which 
        was discovered in 2003 by Brown et al. (2004a) and is often regarded as the key object of this dynamical class (sometimes these 
        objects are called Sednoids). Such spectral differences suggest that Sedna and the other objects do not share the same region of 
        origin (see e.g. Sheppard 2010). This robust observational evidence may be at odds with the Sednitos theory (J{\'{\i}}lkov{\'a} et 
        al. 2015) that argues that Sedna and the other ETNOs were captured from the planetesimal disk of another star early in the history 
        of the Solar System. Sedna may also be a statistical outlier among the ETNOs in terms of some other orbital and physical parameters 
        (de la Fuente Marcos and de la Fuente Marcos 2016c).

        Thanks to the new GTC observations, the orbital solution of 2013~RF$_{98}$ was considerably improved (de Le\'on et al. 2016). Using 
        the new orbits in Table \ref{elements}, the relative differences in (heliocentric) $a$, $e$, $i$, $\Omega$, $\omega$ and time of 
        perihelion passage, $\tau_q$, are respectively 32.6~AU, 0.0465, 3\fdg99, 1\fdg61, 15\fdg26 and 56~days; the associated angular 
        separation between the directions of the perihelia of this pair of ETNOs is now 14\fdg1 (14\fdg1$\pm$0\fdg7) and between the 
        directions of their velocities at perihelion/aphelion is 14\fdg1, but their orbital poles still remain at 4\fdg1 
        (4\fdg059$\pm$0\fdg003) from each other. Having relatively well-aligned orbital poles is indicative of a nearly common direction of 
        orbital angular momentum which is often linked to the products of the break-up of a parent body. For this type of analysis, it may 
        be claimed that when considering objects as distant as the ETNOs, it is perhaps better to use orbital elements that do not vary on 
        very short time-scales, i.e. barycentric instead of heliocentric ones (see e.g. de la Fuente Marcos and de la Fuente Marcos 2016b; 
        Malhotra et al. 2016). However, the positions of orbital poles and perihelia of the ETNOs are fairly insensitive to the differences 
        between heliocentric and barycentric coordinates (de la Fuente Marcos and de la Fuente Marcos 2016c) because these differences are 
        very small (well under 1\%) for the particular case of the angular elements as Table \ref{elements} shows.  
%
%----------------------------------------------------------------------- Orbital elements of the pair of ETNOs (474640) 2004 VN112-2013 RF98 
%
     \begin{table*}
        \fontsize{8}{11pt}\selectfont
        \tabcolsep 0.08truecm
        \caption{Heliocentric and barycentric Keplerian orbital elements of the pair (474640) 2004~VN$_{112}$--2013~RF$_{98}$. In addition 
                 to the nominal value of each parameter, the 1$\sigma$ uncertainty is also indicated. The orbital solutions have been 
                 computed at epoch JD 2457800.5 that corresponds to 00:00:00.000 TDB on 2017 February 16 (J2000.0 ecliptic and equinox. 
                 Source: JPL's Small-Body Database.)
                }
        \centering
        \begin{tabular}{llllll}
           \hline
                                                              &   & \multicolumn{2}{c}{(474640) 2004~VN$_{112}$} & \multicolumn{2}{c}{2013~RF$_{98}$} \\
           \cline{3-6}                                                                                                                                       
                                                              &   & heliocentric       & barycentric             & heliocentric     & barycentric     \\
           \hline
            Semi-major axis, $a$ (AU)                         & = & 316.4$\pm$1.0      & 327.3                   & 349$\pm$11       & 364             \\
            Eccentricity, $e$                                 & = & 0.8505$\pm$0.0005  & 0.8554                  & 0.897$\pm$0.003  & 0.901           \\
            Inclination, $i$ (\degr)                          & = & 25.5848$\pm$0.0002 & 25.5479                 & 29.572$\pm$0.003 & 29.538          \\
            Longitude of the ascending node, $\Omega$ (\degr) & = & 65.9893$\pm$0.0003 & 66.0223                 & 67.596$\pm$0.005 & 67.636          \\
            Argument of perihelion, $\omega$ (\degr)          & = & 327.061$\pm$0.007  & 326.990                 & 311.8$\pm$0.6    & 311.7           \\
            Mean anomaly, $M$ (\degr)                         & = & 0.478$\pm$0.002    & 0.456                   & 0.404$\pm$0.004  & 0.379           \\
            Time of perihelion passage, $\tau_q$ (JED)        & = & 2455069$\pm$2      & 2455064                 & 2455125$\pm$95   & 2455131         \\
                                                              & = & 2009-Aug-25.8      & 2009-Aug-20             & 2009-Oct-20.7    & 2009-Oct-26     \\
            Perihelion, $q$ (AU)                              & = & 47.321$\pm$0.004   & 47.322                  & 36.09$\pm$0.03   & 36.10           \\
            Aphelion, $Q$ (AU)                                & = & 586$\pm$2          & 607                     & 662$\pm$20       & 692             \\
            Absolute magnitude, $H$ (mag)                     & = & 6.5                &                         & 8.7              &                 \\
           \hline
        \end{tabular}
        \label{elements}
     \end{table*}
%
%-------------------------------------------------------------------------------------------------------------------------------------------
%

        At this point it may be argued that a genetic link for the pair of ETNOs 474640--2013~RF$_{98}$ is not sufficiently substantiated,
        that there is a strong observational bias to detecting objects with similar perihelia and poles if discovery surveys are conducted 
        at a similar epoch. It is indeed true that the two ETNOs subject of this investigation have been discovered by the same telescope; 
        however, the instrument was significantly upgraded between 2004 and 2013, and the detector was replaced. The system used in 2013 is 
        not similar to the one used in 2004, and the two surveys were independent, with different pointing strategies (see sect. 2 in de 
        Le\'on et al. 2017). In addition and even if the ETNOs are currently discovered when they are near or at perihelion, their orbital 
        periods are longer than a few thousand years and they spend several decades close to perihelion; therefore, no particularly strong 
        bias towards small relative differences (a few months) in $\tau_q$ is expected. It may also be claimed that no metric is used to 
        confirm that the orbits of the pair of ETNOs 474640--2013~RF$_{98}$ are dynamically similar. These metrics are customarily applied 
        when defining asteroid family links within the main asteroid belt (see e.g. Milani et al. 2014; Nesvorn{\'y} et al. 2015), but these 
        asteroid families are mostly collisional in origin while the pair studied here could be the result of tidal stripping induced during 
        a planetary encounter. In any case, if such metrics are applied to the current sample of known ETNOs the values obtained are all 
        well above the thresholds used to define asteroid families in the main belt. 

     \subsection{Pole and perihelion separations: out of the ordinary or not?}
        Within the context of the standard eight-planets-only Solar System paradigm, the distributions of the orbital parameters of minor 
        bodies following orbits similar to the ones of the ETNOs should be statistically compatible with those of an unperturbed asteroid 
        population moving in heliocentric Keplerian orbits (particularly in the case of objects like Sedna or 2012~VP$_{113}$). Assuming 
        such an unperturbed scenario and using a model similar to the one described by de la Fuente Marcos and de la Fuente Marcos (2014), 
        the probability of finding values of the angular separations of the pertinent directions as low as those of the pair of ETNOs 
        474640--2013~RF$_{98}$ (see above) by chance is less than 0.0019. 
  
        This value of the probability has been computed using a Monte Carlo approach that generates a synthetic population of ETNOs with 
        $a\in$ (150, 800) AU, $e\in$ (0.70, 0.95), $i\in$ (0, 55)\degr, $\Omega\in$ (0, 360)\degr, and $\omega\in$ (0, 360)\degr, assuming 
        that the orbits are uniformly distributed in orbital parameter space. These ranges are the same ones used in the recent ETNO 
        detectability study carried out by Shankman et al. (2017b). Although out of the scope of this paper, our approach can reproduce 
        reasonably well most of the features present in figs. 1 and 2 of Shankman et al. (2017b). We restrict the analysis to virtual 
        objects with perigee $<90$~AU. From this synthetic population, we single out those objects with $q>$ 30 AU and $|\delta|<24${\degr}, 
        about 20\% out of 10$^{7}$ pairs. For two such random orbits, the probability of the perihelion directions being within 15{\degr} 
        of each other and, concurrently, the pole directions being closer than 5{\degr} has been evaluated in the usual way, counting the 
        number of relevant pairs and dividing by the total number (see e.g. Wall and Jenkins 2012). If $q>$ 40~AU is selected instead of 
        $q>$ 30 AU, the value of the probability is still less than 0.0019. Although the input distributions in $\Omega$ and $\omega$ are 
        uniform in the interval (0,~360)\degr, the resulting output distributions ---obtained after imposing the various constraints--- used 
        to compute the angular separations are very different from the input ones (see fig. 3 in de la Fuente Marcos and de la Fuente Marcos 
        2014).

        In this analysis, the probabilities associated with perihelia and poles are not independent as the location of these points is 
        computed using expressions that share one or more orbital parameters (de la Fuente Marcos and de la Fuente Marcos 2016c). Our Monte 
        Carlo analysis shows that the probability of two objects having an angular separation between the directions of their perihelia 
        $<15${\degr} is 0.030; a similar calculation performed to find the probability of having an orbital pole separation $<5${\degr} 
        gives a value of 0.023. The incorrect assumption of treating them as independent would lead us to evaluate the probability of 
        interest here as simply the product of probabilities (or less than 0.0007), which is wrong. In Batygin and Brown (2016), when 
        studying the issue of clustered perihelia and orbital pole positions, it is assumed that the two measurements are statistically 
        uncorrelated and the joint probability of observing both clustering in perihelion position and in pole orientation concurrently is 
        found multiplying the probabilities together.

        A somewhat similar study, but focusing on the putative clustering of poles and perihelia of the orbits of long-period comets, was 
        carried out by Bogart and Noerdlinger (1982). Their investigation was limited to the three elements that specify the spatial 
        orientation of an orbit ($i$, $\Omega$ and $\omega$); in contrast, our Monte Carlo approach includes all the orbital parameters. If 
        we apply eq. (1b) in Bogart and Noerdlinger (1982) considering ($N=$) 28 random orbits, a separation in orbital plane normals ($X=$) 
        of 5\degr, and a separation in perihelion directions ($Y=$) of 15\degr, we obtain an average number of pairs expected within those 
        ranges of relevant angular separations of 0.061 or a probability of 0.00016. Therefore, finding one pair like 474640--2013~RF$_{98}$ 
        is unlikely assuming that the ETNOs constitute an unperturbed asteroid population, but we do not know whether this pair is a true 
        outlier or the ETNOs are indeed a perturbed population.

        The value obtained when we apply eq. (1b) in Bogart and Noerdlinger (1982) is lower than our own, which indicates that using the 
        entire orbit and restricting the values of $q$ plays a role, as the size and shape of the orbits were neglected in their work. 
        However, it still conveys the same message, that the existence of the ETNO pair 474640--2013~RF$_{98}$ is probably not compatible 
        with an unperturbed scenario or chance. As the visible spectra of the members of the pair are also close matches, a putative common 
        physical origin is likely. As of the time of writing, only Sedna, 474640 and 2013~RF$_{98}$ have spectroscopic results published. 
        Without compositional information it is not possible to argue for a common genetic origin, no matter how similar the orbits of the 
        objects involved are. In this respect, the pair 474640--2013~RF$_{98}$ is unique at the moment. Viable scenarios that could explain 
        these results include fragmentation and binary dissociation. In both cases, the presence of an unseen massive perturber, i.e. a 
        trans-Plutonian planet, may be required. 

     \subsection{Fragmentation versus binary dissociation: where are the binaries?}
        Close encounters between minor bodies and planets can induce fragmentation (e.g. Scheeres et al. 2000; Sharma et al. 2006; Ortiz et 
        al. 2012), but the minimum approach distance associated with such events (about 20 planetary radii, see e.g. Keane and Matsuyama 
        2015) is far shorter than the one required for binary dissociation in the case of wide binaries whose binding energies are rather 
        small (Agnor and Hamilton 2006; Vokrouhlick\'y et al. 2008; Parker and Kavelaars 2010). The simplest formation mechanism for 
        dynamically-related asteroid pairs is binary disruption during planetary flybys, but it requires the presence of binary asteroids 
        moving on planet crossing orbits (e.g. Jacobson 2016). How likely is this possibility in our case? 

        The existence of wide binaries among the asteroid populations orbiting beyond Neptune is a well-documented fact (e.g. Parker et 
        al. 2011); the widest known binary is 2001~QW$_{322}$ with a binary semi-major axis of 102\,100~km (Petit et al. 2008; Parker et 
        al. 2011). They have been found preferentially among the dynamically cold, classical TNOs, but they are present within the scattered 
        TNO population as well.\footnote{\url{http://www2.lowell.edu/~grundy/tnbs/status.html}} Dysnomia, the satellite of the dwarf planet 
        Eris, revolves about its host with a binary semi-major axis of 37\,400~km (Brown et al. 2005, 2006). The moon of (225088) 
        2007~OR$_{10}$ may have a binary semi-major axis of 29\,300~km (Kiss et al. 2017). The only known satellite of the dwarf planet 
        Makemake has a binary semi-major axis of over 21\,000~km (Parker et al. 2016). The satellite of 2004~PB$_{108}$ has a binary 
        semi-major axis of 10\,400~km (Grundy et al. 2009). 

        All TNOs larger than about 1\,000~km are known to harbour one or more moons (Barr and Schwamb 2016; Kiss et al. 2017). Fraser et al. 
        (2017a, 2017b) have found that the blue-coloured (spectral slope $<17$\%), cold TNOs are predominantly in tenuously bound binaries 
        and proposed that they were all born as binaries at $\sim$38~AU. Previous studies had estimated that the binary fraction among the 
        dynamically cold TNOs could be about 30\% (Grundy et al. 2011) and it might reach $\sim$10\% for the dynamically excited populations 
        (Noll et al. 2008). It is thought that 10\% to 20\% of all TNOs could host one or more gravitationally bound companions (Brown et 
        al. 2006), but see the detailed discussion in Petit and Mousis (2004). 

        Binary ETNOs have not yet been discovered, but Sedna (Brown et al. 2004b) and 474640 (Fraser and Brown 2012) have been observed with 
        the Hubble Space Telescope and no close companions have been reported yet. In addition, most ETNOs are perhaps too faint to be 
        observed by adaptive optics on even the largest existing ground-based telescopes, although detection biases favour the discovery of 
        wide binaries. 

     \subsection{Summary: a reasonably sound scientific case}
        Although the hypothesis of production of dynamically coherent pairs of ETNOs by binary dissociation induced by close encounters with 
        a massive planetary perturber is interesting in its own right, it may be argued that this scenario appears to be very unlikely, as 
        it requires a number of concurrent, low-probability ingredients in order to make it work. 

\begin{figure}
        \centering
        \includegraphics[width=\linewidth]{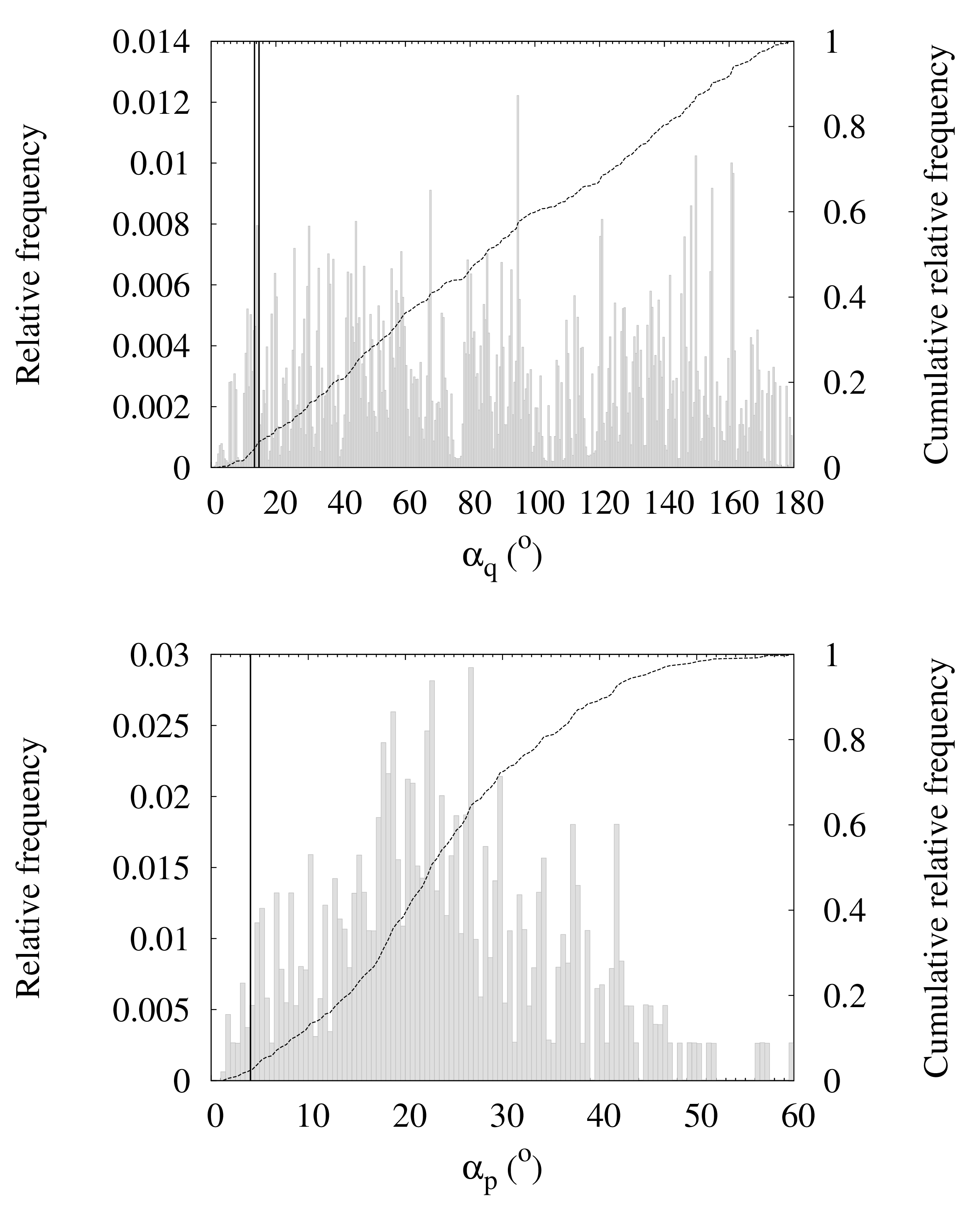}
        \caption{Possible distributions (the bin size is 0\fdg5) of the angular separations between orbital poles, $\alpha_{\rm p}$, and 
                 perihelia, $\alpha_q$, for the known ETNOs. These are the result of the analysis of 10$^{7}$ random pairs of ETNOs with 
                 synthetic orbits based on the mean values and dispersions of the orbital elements of real ETNOs (see the text for details). 
                 The observed dispersion (1$\sigma$) of the values for the ETNO pair 474640--2013~RF$_{98}$ is represented by vertical 
                 lines; this is clearly not an ordinary pair of ETNOs.
                }
        \label{angdistri}
     \end{figure}

        First, one may argue that the ETNO pair 474640--2013~RF$_{98}$ is just one out of many similar pairs of ETNOs. However, this 
        statement is far from true. With 28 known ETNOs, we now have 378 different pairs. The pair studied in this work is one of only two 
        with a separation between orbital poles $<5${\degr} and a gap between directions of perihelia $<15${\degr}; its associated $p$-value 
        is therefore equal to 0.00529. If 474640--2013~RF$_{98}$ is an ordinary pair of ETNOs (from the point of view of the orientations in 
        space of the orbits of both components), the probability of observing another pair of ETNOs with relevant angular separations as 
        small or smaller than those of this pair should be relatively large. In striking contrast, a low value of the probability is found 
        that we interpret as strong evidence against 474640--2013~RF$_{98}$ being a regular pair of ETNOs. It can still be argued that the 
        orbital elements of the ETNOs are affected by uncertainties that a simple computation of probabilities like the previous one cannot 
        take into account; therefore, its results cannot be relied upon at all. Following the ideas discussed by e.g. Fisher (1935), Basu 
        (1980) and Welch (1990), we have used the available data provided by Jet Propulsion Laboratory (JPL)'s Solar System Dynamics Group 
        Small-Body Database (Giorgini 2011, 2015)\footnote{\url{https://ssd.jpl.nasa.gov/sbdb.cgi}} to generate 10$^{7}$ random pairs of 
        virtual ETNOs and computed the angular separation between orbital poles, $\alpha_{\rm p}$, and perihelia, $\alpha_q$, of each of 
        them. The angles have been calculated as described by de la Fuente Marcos and de la Fuente Marcos (2016c). Regarding the computation
        of the individual random orbits and focusing e.g. on the inclination parameter, a new value has been found using the expression 
        $i_{\rm r} = \langle{i}\rangle + \sigma_{i}\,r_{\rm i}$, where $i_{\rm r}$ is the inclination of the random orbit, 
        $\langle{i}\rangle$ is the mean value of the inclination of one of the real ETNOs, $\sigma_{i}$ is its associated standard 
        deviation, and $r_{\rm i}$ is a (pseudo) random number with normal distribution in the range $-$1 to 1. The resulting distributions 
        of possible relevant angular separations of the ETNOs are plotted in Fig.~\ref{angdistri}; the actual dispersion in the values of 
        the observed angles for the ETNO pair 474640--2013~RF$_{98}$ is represented by vertical lines. From this analysis, the probability 
        of finding a pair of ETNOs with values of $\alpha_{\rm p}$ and $\alpha_q$ both below those of the pair 474640--2013~RF$_{98}$ 
        (4\fdg1 and 14\fdg1, respectively) is 0.00449$\pm$0.00002 (average and standard deviation from 10 sets of experiments). This value 
        is independent of any assumptions made regarding the distributions and ranges of the various orbital elements (as we did in 
        Sect.~2.2) and it is far too small to make this pair of ETNOs representative of the typical behaviour, in terms of relative orbital 
        orientation in space, of the pairs present in the sample of known ETNOs.
%
%-------------------------------------------------------------------------------------------------------------------------------------------
%
%
%-------------------------------------------------------------------------------------------------------------------------------------------
%

        Second, it can be claimed that there are no scientific grounds to argue for the presence of an unknown perturber, massive enough to
        cause binary dissociations and orbiting the Sun at hundreds of astronomical units. On the one hand, any unknown planet located 
        between the trans-Neptunian belt (see e.g. Fern\'andez 1980; Jewitt and Luu 1993) and the Oort Cloud (see e.g. Oort 1950) must have 
        a mass less than that of Saturn ($5.68319 \times 10^{26}$~kg or over 95 Earth masses) to have escaped detection by the all-sky 
        {\it WISE} survey (Luhman 2014); on the other hand, Larsen et al. (2007) have shown that their Spacewatch results obtained within 
        10{\degr} of the ecliptic exclude the presence of any Mars-sized objects out to 300~AU and Jupiter-sized planets out to 1,200~AU. 
        While the {\it WISE} survey was all-sky, the Spacewatch project covered 8,000 square degrees, avoiding the regions towards the 
        Galactic centre (the clouds of Sagittarius and their neighbourhood), but focusing on low inclinations. Lykawka and Mukai (2008) have 
        suggested that a planetary body smaller than the Earth could be following an eccentric and inclined orbit between 100 and 200~AU 
        from the Sun, and cause the so-called Kuiper Cliff (see e.g. Chiang and Brown 1999). Holman and Payne (2016a) studied the available 
        astrometry of Pluto and other TNOs to conclude that the presence of a planet at 60--100~AU with a mass as low as 0.6--3 Earth masses 
        ---from their eq. (5)--- could not be ruled out. Volk and Malhotra (2017) have found robust statistical evidence that the mean plane 
        of the trans-Neptunian belt is warped in such a way that an inclined, low-mass (probably Mars-sized), unseen planet could be 
        responsible for the warping. In any case and although it may have an apparent magnitude in excess of 18, the presence of such a 
        perturber (up to a few Earth masses) has not been effectively rejected by past or present surveys, particularly if it moves in a 
        moderately inclined orbit (above 10\degr) and/or close to the Galactic plane (see e.g. Larsen et al. 2007; Brown et al. 2015). The 
        possible perturber considered in our work is more distant and more massive than the one thought to be sculpting the Kuiper Cliff, 
        but its existence has not been ruled out by recent analyses carried out by e.g. Brown and Batygin (2016), Fienga et al. (2016), 
        Holman and Payne (2016a, 2016b), Sheppard and Trujillo (2016), and Brown (2017). In addition, the study of exo-planetary systems 
        shows that planets moving in very wide orbits indeed exist (see e.g. Bailey et al. 2014; Naud et al. 2014) and that most exo-planets 
        have values of their masses below those of Uranus and Neptune but above that of the Earth (see e.g. Howard et al. 2010; Malhotra 
        2015; Silburt et al. 2015).

        Third, if the pair of ETNOs is indeed unusual and (at least) one yet-to-be-detected distant planetary-mass perturber goes around the 
        Sun between the trans-Neptunian belt and the Oort Cloud, how a wide binary asteroid may have survived as a bound pair until the 
        relatively recent past? The answer to this legitimate question is not an easy and straightforward one, mainly because we do not know 
        yet the actual source or sources of this population. If the source of the ETNOs is in the Oort Cloud, it is unclear what is the 
        binary fraction there because no binary comets have ever been observed, although some can be considered as contact-binary comets 
        ---see e.g. 8P/Tuttle that has a strongly bifurcated nucleus (Harmon et al. 2010) or 67P/Churyumov-Gerasimenko (Sierks et al. 2015); 
        the same can be said about a source within the inner Oort Cloud (Hills 1981). An origin in the region of the Giant planets early in 
        the history of the Solar System can be readily discarded because a loosely-bound pair could not possibly survive recurrent 
        gravitational encounters with the Jovian planets. As pointed out above, a possible source for tenuously bound TNO binaries has been 
        identified by Fraser et al. (2017a, 2017b) at $\sim$38~AU and one may speculate that a similar source may exist well beyond the 
        trans-Neptunian belt. In any case, if planets can form at 125--750~AU from the Sun (Kenyon and Bromley 2015, 2016), it is difficult 
        to argue that minor bodies (and perhaps binaries) cannot.

        Fourth, the frequency of such encounters, assuming that the perturber (the trans-Plutonian planet) and the target (the wide binary 
        asteroid) do exist, is also a matter of concern. The classical method of \"Opik (1951) and Wetherill (1967) has been recently 
        revisited by JeongAhn and Malhotra (2017). An application of this theory results in an average value of the collision probability per 
        year and pair of objects of the order of $10^{-10}$. Considering the age of the Solar System, 4,500 Myr, and that the inner Oort 
        Cloud may have millions of members (Hills 1981; Levison et al. 2001), the existence of a non-negligible number of pairs of 
        present-day ETNOs resulting from the dissociation of wide binaries is entirely possible.

        In conclusion, based on our probabilistic argument, it is rather difficult to argue that the ETNO pair 474640--2013~RF$_{98}$ is 
        just a standard couple of ETNOs; this pair is a true outlier. The orientations of their orbits are simply too well correlated to be 
        the result of chance alone. Having correlated orientations implies a level of dynamical coherence only attainable as a result of 
        fragmentation processes or binary dissociation. Fission could be the result of fast rotation, internal processes, or tidal 
        encounters with massive perturbers. There are no currently known candidate mechanisms able to induce fast rotation or spontaneous 
        fragmentation of minor bodies at hundreds of astronomical units from the Sun. In addition, fragmentation via tidal encounters is far 
        less probable than binary dissociation as the encounters must take place at a much closer range in the first case. Therefore, we 
        arrive at our proposed scenario by discarding other options that, in principle, appear to be far less probable and much more 
        speculative. But, within the trans-Plutonian planets paradigm and assuming that wide binaries are also present among the ETNO
        population, how are they affected by interactions with a planet?

  \section{Dissociation of wide binary ETNOs: an \textit{N}-body approach}
     In order to study the dissociation of wide binary ETNOs during close encounters with hypothetical trans-Plutonian planets, we use 
     direct $N$-body simulations performed with a modified version of a code written by Aarseth (2003) that implements the Hermite 
     integration scheme described by Makino (1991) as a fourth-order method. The standard version of this code is publicly available from 
     the IoA website;\footnote{\url{http://www.ast.cam.ac.uk/~sverre/web/pages/nbody.htm}} the version used in this research includes 
     purpose-specific input/output modifications. The value of the dimensionless time-step factor {\it ETA} was very conservative, leading
     to very low typical relative energy errors; the total relative error in the value of the energy at the end of our integrations was 
     always $\leq10^{-12}$ and often as low as $5\times10^{-14}$. The quality of the results obtained with this software applied to Solar 
     System calculations has been positively and extensively assessed by de la Fuente Marcos and de la Fuente Marcos (2012); in particular, 
     fig. 3 in de la Fuente Marcos and de la Fuente Marcos (2012) shows that the results of long-term integrations performed with the 
     program used in this study are similar to those obtained with other, well-tested codes.

     Following the analysis by de la Fuente Marcos et al. (2016), our physical model includes the perturbations by the Jovian planets 
     (Jupiter to Neptune) and one trans-Plutonian planet. Initial positions and velocities of both known planets and ETNOs are based on the 
     DE405 planetary orbital ephemerides (Standish 1998) referred to the barycentre of the Solar System and to the epoch JD TDB 2457800.5 
     (2017-February-16.0), which is the $t$ = 0 instant in our calculations. Heliocentric and barycentric Keplerian orbital elements of the 
     pair (474640) 2004~VN$_{112}$--2013~RF$_{98}$ (see Table \ref{elements}) were provided by JPL's On-line 
     Solar System Data Service\footnote{\url{http://ssd.jpl.nasa.gov/?planet_pos}} (Giorgini et al. 1996). Orbital elements are transformed 
     into barycentric initial positions and velocities as needed. Two types of numerical experiments have been performed using the same 
     software and physical model. 

     The first one (Sect. 4) is designed to explore the binary dissociation process itself as wide binary ETNOs experience encounters with 
     one trans-Plutonian planet. Binary destruction could be the result of the total energy of the system becoming greater than zero, but 
     also of enlargement of the binary semi-major axis beyond one Hill radius with respect to the Sun or even a collision. Due to the large 
     size and high eccentricity of the orbits of the ETNOs, in this work we focus on physically unbound systems (energy condition) but 
     (binary semi-major axis) enlargement-driven dissociations are also identified. The second type (Sect. 5) applies integrations 
     backwards in time (see also de Le\'on et al. 2017), beginning with the present-day orbits of the unbound pair 474640--2013~RF$_{98}$ 
     (see Table \ref{elements}), to investigate what properties a perturber should have to induce the observed tilt between the orbital 
     planes of these ETNOs (4\fdg1) starting from the values characteristic of newly disrupted pairs found from the analysis of the first 
     set of numerical experiments.

\begin{figure}
        \centering
        \includegraphics[width=\linewidth]{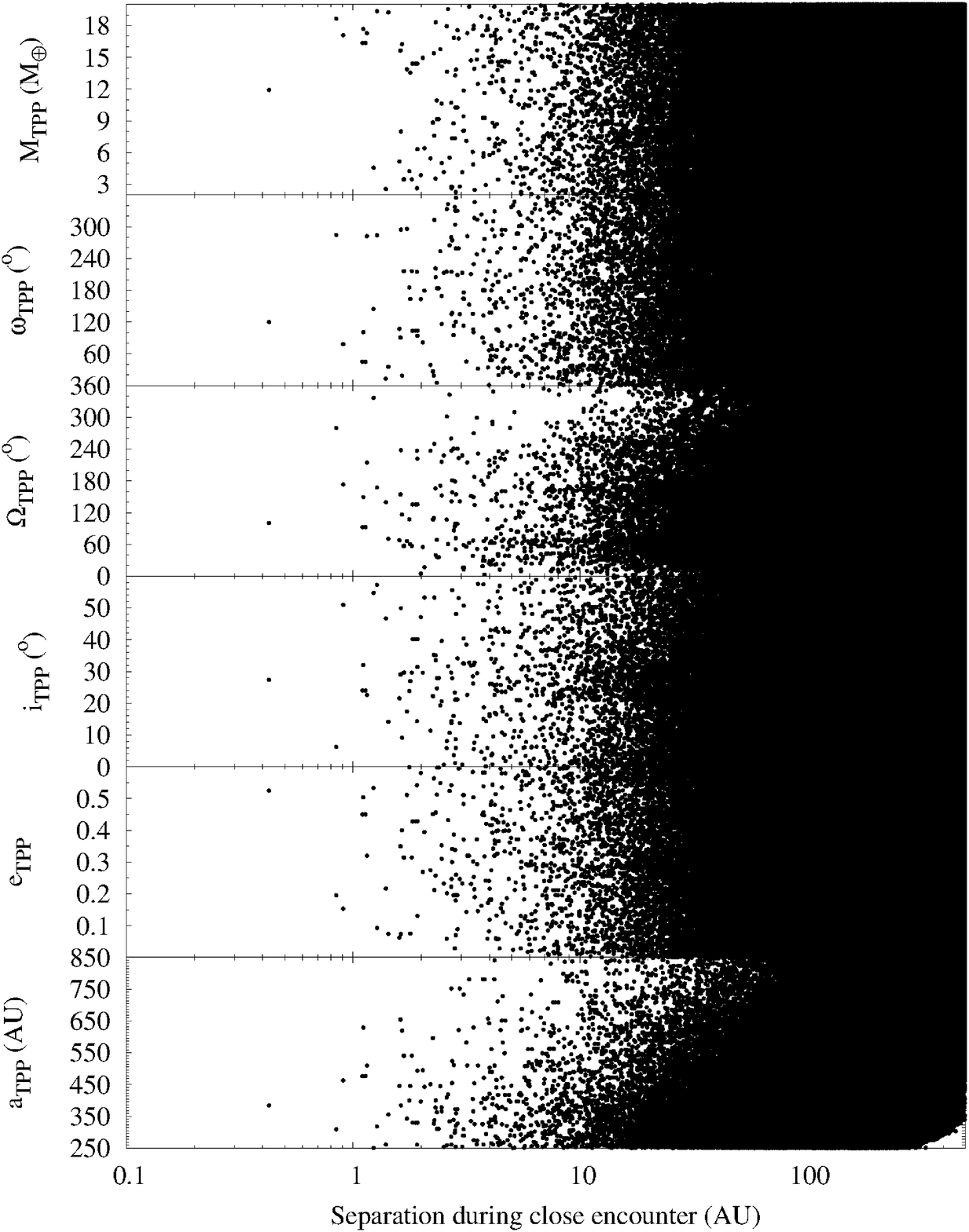}
        \caption{Orbital elements and mass of a sample of trans-Plutonian planets (TPPs) undergoing close encounters with a wide binary 
                 ETNO moving in an orbit compatible with that of (474640) 2004~VN$_{112}$. The x-axis shows the minimum separation between 
                 binary and planet during the simulated close encounter. The solution associated with the closest approach was further 
                 refined to perform the experiments whose results are reported in Sect. 4. The results of 10$^{6}$ experiments are 
                 plotted.
                }
        \label{close}
     \end{figure}

     For those experiments involving binaries and in order to compute the test orbit of the centre of masses of the binary, we consider how
     the elements influence each other and their associated uncertainties, applying the implementation of the Monte Carlo using the 
     Covariance Matrix (MCCM) method discussed by de la Fuente Marcos and de la Fuente Marcos (2015). The binary orbits studied here have 
     initial parameters drawn from a nominal orbit adding random noise to each initial orbital element as described by the covariance 
     matrix. The covariance matrices used here were provided by JPL's Small-Body Database\footnote{\url{http://ssd.jpl.nasa.gov/sbdb.cgi}} 
     and the vector including the mean values of the orbital parameters at the given epoch is of the form $\textit{\textbf{v}} = (e, q, 
     \tau_{q}, \Omega, \omega, i)$. Suitable initial conditions for the trans-Plutonian planet included in the simulations that result in 
     binary dissociation events (Sect. 4) were identified by performing a preliminary experiment to single out candidate solutions of 
     perturbers that may pass close enough to the binary ($<2$~AU) for an integrated time of 8\,000 yr. The minimum separations between 
     binary and planet during the simulated close encounters resulting from this experiment are plotted in Fig. \ref{close}. The input 
     ranges for the parameters of the perturber (orbital elements and mass) were obtained from the set of experiments discussed in Sect. 5. 
%
%-------------------------------------------------------------------------------------------------------------------------------------------
%
%
%-------------------------------------------------------------------------------------------------------------------------------------------
%

  \section{From wide binary ETNO to the newly disrupted state}
     The numerical experiments described in this section include a binary that follows a heliocentric orbit consistent with that of the 
     present-day ETNO (474640) 2004~VN$_{112}$ (see Table \ref{elements}); the binary experiences a flyby with a planetary perturber at 
     hundreds of astronomical units from the Sun. This is a somewhat arbitrary but reasonable choice because the main objective of this 
     study is neither reconstructing in detail the past dynamical history of the pair of ETNOs 474640--2013~RF$_{98}$ nor making an 
     exhaustive exploration of the dynamical pathways leading to present-day pairs of ETNOs, but showing the feasibility of the 
     binary-planet interaction scenario as a source of related pairs of ETNOs. 

     The heliocentric orbit of the binary at the beginning of each experiment is computed using the MCCM method (see above). The masses of 
     the binary components are assumed to be 2.1$\times$10$^{19}$ kg for the primary and 1.0$\times$10$^{18}$ kg for the secondary; these 
     values are consistent with results obtained by Parker et al. (2011) and de Le\'on et al. (2017). The orbital elements of the binary 
     are drawn from uniform distributions with ranges $a_{\rm b}\in(10\,000, 400\,000$)~km, $e_{\rm b}\in(0.1, 0.9)$ but imposing a starting 
     value of the binary apocentre $<600\,000$~km to ensure initial stability, $i_{\rm b}\in(0, 180)$\degr, $\Omega_{\rm b}\in(0, 360)$\degr, 
     and $\omega_{\rm b}\in(0, 360)$\degr. For computational convenience, the binaries are always started at apocentre. The initialization 
     of the binary components is carried out as described in sect. 8.3 of Aarseth (2003). 

     The upper limit in $a_{\rm b}$ is somewhat arbitrary as no binary ETNOs have been detected yet, but the widest binary TNOs have average
     separations in units of the radius of the primary of the system $\leq1\,000$, with 2001~QW$_{322}$ being an outlier at 2\,200 (Petit et 
     al. 2008). Although most separations are $\leq$600, 2000~CF$_{105}$ (Noll et al. 2002; Parker et al. 2011), 2003~UN$_{284}$ (Millis and 
     Clancy 2003; Parker et al. 2011) and 2005~EO$_{304}$ (Kern et al. 2006; Parker et al. 2011) have values close to or slightly above 
     1\,000. All these objects are cubewanos. Within this context and assuming primaries with sizes in the range 300--400~km, a value of the 
     binary semi-major axis under 400\,000~km does not seem implausible.  
 
\begin{figure}
        \centering
        \includegraphics[width=\linewidth]{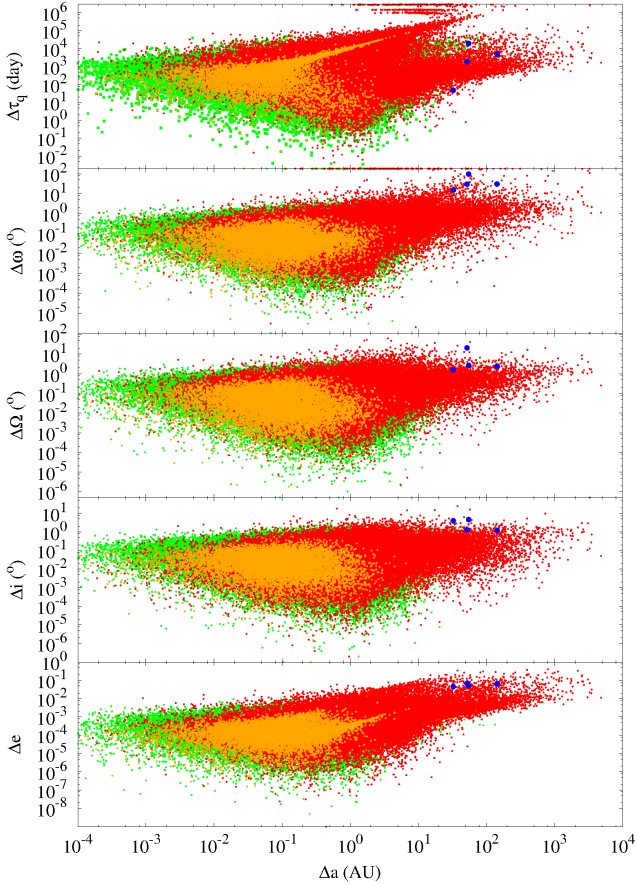}
        \caption{Differences between the values of the heliocentric orbital elements of the members of the initially bound binary at the
                 end of the simulation (8\,000 yr). Bound pairs are plotted in green, unbound couples in red (energy condition) and orange 
                 (separation condition); the blue points show the actual values for the pairs of ETNOs (474640) 
                 2004~VN$_{112}$--2013~RF$_{98}$ ($\Delta{a}=32.6$~AU), 2002~GB$_{32}$--2003~HB$_{57}$ ($\Delta{a}=51.9$~AU), (82158) 
                 2001~FP$_{185}$--2013~UH$_{15}$ ($\Delta{a}=54.8$~AU), and (148209) 2000~CR$_{105}$--2010~GB$_{174}$ 
                 ($\Delta{a}=143.1$~AU). These pairs have angular separations of the orbital poles $<10$\degr. The results of 500\,000 
                 experiments are plotted, excluding hyperbolic ejections and collisions/captures.
                }
        \label{BiDi}
     \end{figure}

     The orbit of the perturber ---$a=399.61\pm0.06$~AU, $e=0.307\pm0.002$, $i=23.86\pm0.03$\degr, $\Omega=77.76\pm0.05$\degr, 
     $\omega=35.06\pm0.05$\degr, and true anomaly, $f=19.07\pm0.07$\degr--- is based on a refinement of the optimal candidate 
     orbits resulting from the experiment plotted in Fig. \ref{close}; its randomized mass is assumed to be in the range 
     2--20~$M_{\oplus}$. With such orbit and range of masses, the value of the Hill radius of the perturber is in the range 3.49--7.52~AU;
     relevant minimum separations between binary and perturber during close encounters are well below this range of values. The encounters 
     take place at 327.5$\pm$0.9~AU from the Sun. For each experiment, the orbit of the perturber is drawn from uniform distributions 
     defined by the assumed ranges. In order to minimize the chances of a physical collision at the binary ETNO or a capture as satellite by 
     the planet, we perform hundreds of thousands of short (8\,000 yr or nearly one orbital period of the perturber) eight-body simulations. 
     The results of these short numerical experiments are fully consistent with those of longer ones (24\,000 yr, compare Figs. \ref{BiDi} 
     and \ref{BiDi++}) for the relevant section of the relative orbital parameter space ($\Delta{a}>10$~AU, see below); however, the overall 
     fraction of unbound pairs increases by 1\% when the time interval is tripled. This arbitrary choice for the duration (24\,000 yr) of 
     this second set of control calculations has nothing to do with the existence of a time window of observability of the members of the 
     pair after disruption. The sole purpose of this additional set of control calculations is gaining a better understanding of the 
     border-line cases of binary dissociation, but these cases are not central to our study. 

     Figure \ref{BiDi} shows the differences between the values of the heliocentric orbital elements of the members of the initially bound 
     binary at the end of the simulation for 500\,000 experiments (differences for angular elements $\leq180$\degr). Pairs that are 
     still bound (energy condition) at the end of the simulation are plotted in green, unbound couples in red if the total energy of the 
     system is greater than zero or orange if the binary semi-major axis is greater than one Hill radius of the primary (which is of order 
     of 10$^{6}$~km or about 0.007~AU in our case) but the relative energy is still negative; the pairs of ETNOs 474640--2013~RF$_{98}$, 
     2002~GB$_{32}$--2003~HB$_{57}$, (82158) 2001~FP$_{185}$--2013~UH$_{15}$, and (148209) 2000~CR$_{105}$--2010~GB$_{174}$ (sorted by 
     increasing $\Delta{a}$), are plotted in blue for comparison. These pairs are the ones with the lowest values of the angular separation 
     of the orbital poles of their components ($<10${\degr}, de la Fuente Marcos and de la Fuente Marcos 2016c) and, therefore, the most 
     probable by-products of the dynamical scenario under study here. 

     Out of 500\,000 experiments, 22.3\% (111\,307) produced a newly disrupted couple. Out of these unbound pairs, 87.9\% (97\,851) were
     physically unbound systems, the remaining 12.1\% (13\,456) experienced mutual orbit expansion beyond one Hill radius. The fraction of 
     collisions and captures was 1.3\% (6\,490). Nearly 1.3\% (6\,427) of the experiments performed resulted in the hyperbolic ejection of 
     at least one member of the pair, i.e. one or both components escaped from the Solar System to become interstellar minor 
     bodies.\footnote{For the numerical experiments lasting 24\,000 yr, Fig. \ref{BiDi++}, 23.3\% (116\,561) resulted in binary 
     dissociation, out of these, 95.0\% (110\,752) had positive relative energy; the fraction of ejections/collisions/captures was 1.3\%.} 
     Out of the full pairs that left the Solar System, about 29.6\% (1\,869) were still bound as binaries. Over 0.03\% of the experiments 
     produced a disrupted couple with just one member being ejected from the Solar System. Figures \ref{BiDi} and \ref{BiDi++} show that the 
     fraction of disrupted couples with total energy of the system still negative decreases significantly over time (from 12.1\% to 5\% as 
     they become physically unbound) and also that the most sensitive parameter regarding time evolution is $\tau_q$ (compare top panels). 
     The fraction of unbound pairs with $\Delta{a}>10$~AU and difference in $\tau_q$ shorter than one year becomes practically negligible 
     after 24\,000 yr of evolution; i.e. if $\Delta\tau_q$ is very short (weeks to months), the unbound pair must be dynamically very young 
     or the short $\Delta\tau_q$ must be due to chance.

     Our results show that, within the scenario discussed here, wide binary ETNOs can fully dissociate during close encounters with 
     hypothetical trans-Plutonian planets and most binaries are destroyed when the system becomes physically unbound. However, the majority 
     of unbound pairs with $\Delta{a}<10$~AU are borderline cases where the total energy of the binary system is positive, but only by a 
     slight margin. The total energy of these systems may become marginally negative at a later time. The fraction of these systems is 
     somewhat reduced in longer integrations as they become more tightly bound (or the components recede further from each other) over time 
     (see Fig. \ref{BiDi++}). Unbound pairs with larger $\Delta{a}$ are {\it bona fide} newly disrupted couples; these are the ones of 
     interest here and their relative numbers do not change significantly between Figs. \ref{BiDi} and \ref{BiDi++} (1.8\% versus 2.2\%).
%
%-------------------------------------------------------------------------------------------------------------------------------------------
%
%
%-------------------------------------------------------------------------------------------------------------------------------------------
%

     Figure \ref{disini} shows the frequency distribution of the initial orbital elements of the binaries that became unbound pairs. Wider 
     pairs ($a_{\rm b}>0.0015$~AU, see Fig. \ref{disini}, top panel) are far more likely to suffer dissociation (about 5 times) than 
     tighter ones, but the role of the eccentricity, inclination, longitude of the ascending node, and argument of pericentre of the binary 
     on the overall disruption results is minor (see Fig. \ref{disini}, second panel and below). More eccentric binaries are a bit more 
     vulnerable to disruption, as are those with $i_{\rm b}>40$\degr; binaries with $\Omega_{\rm b}$ close to 140{\degr} or 320{\degr} 
     appear to a certain extent less prone to become disrupted couples. If $a_{\rm b}\geq0.002$~AU, the fraction of binary disruptions is 
     close to 50\%. For $a_{\rm b}\sim$0.0015~AU, nearly 25\% of the initially bound binaries are disrupted. If we focus on binaries with 
     $a_{\rm b}<150\,000$~km (or 0.001~AU), the fraction of unbound couples at the end of the simulation is $<15$\%. For this tighter group, 
     the role of the binary orbital elements (other than $a_{\rm b}$) on the disruption outcome is in every way negligible.

     Figure \ref{approdis}, top panel, shows that most binary dissociation events are the result of close encounters at minimum approach 
     distances smaller than about 0.25~AU. For encounters under 1~AU over 20\% of the binaries become unbound pairs; however, nearly 75\% of 
     the binaries reaching separations from the perturber below 0.1~AU are disrupted. This is to be expected because if the planetocentric 
     trajectory of the binary is hyperbolic, the deeper the encounter, the stronger its effects on the orbital parameters of the binary. 
     For a given minimum approach distance, the binary destruction fraction depends on the value of $a_{\rm b}$. However, the outcome of 
     close encounters under 0.01~AU is virtually insensitive to the value of the binary semi-major axis. If we consider binaries with 
     $a_{\rm b}<150\,000$~km that approach the perturber inside 0.01~AU, nearly 75\% of them are disrupted; in stark contrast, just about 
     6\% of them become unbound if the encounter is below 1~AU. Nearly 10\% of the unbound pairs (11\,442 out of 111\,307) had 
     $a_{\rm b}<150\,000$~km. In general, most unbound pairs with $\Delta{a}>10$~AU are the result of close encounters inside 0.1~AU that is 
     about 25 times the value of the maximum initial binary apocentre ($600\,000$~km). 
%
%-------------------------------------------------------------------------------------------------------------------------------------------
%
     \begin{figure}
        \centering
        \includegraphics[width=\linewidth]{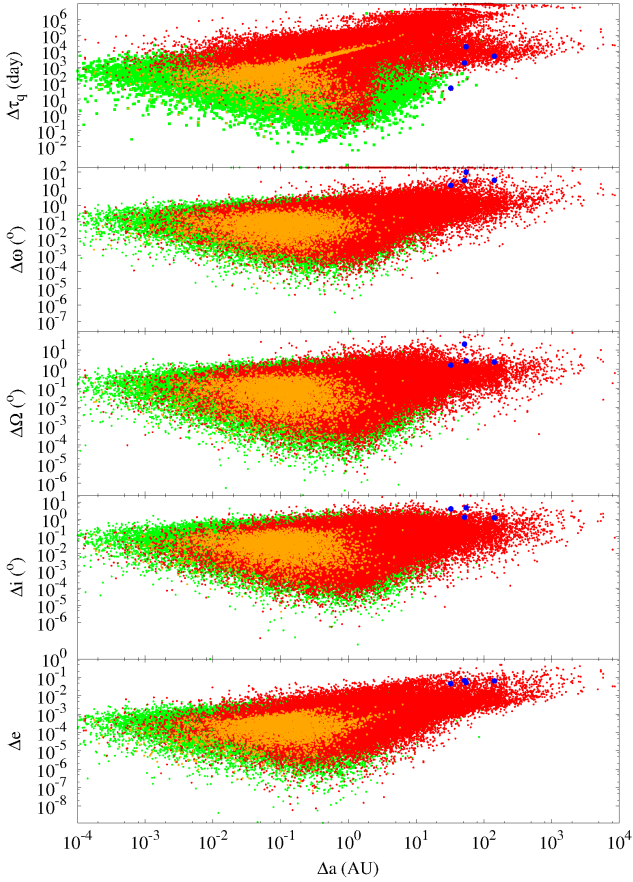}
        \caption{As Fig. \ref{BiDi} but for integrations lasting 24\,000 yr instead of 8\,000 yr. The results of 500\,000 experiments are 
                 plotted, excluding hyperbolic ejections and collisions/captures.
                }
        \label{BiDi++}
     \end{figure}
%
%-------------------------------------------------------------------------------------------------------------------------------------------
%
%
%-------------------------------------------------------------------------------------------------------------------------------------------
%
     \begin{figure}
        \centering
        \includegraphics[width=\linewidth]{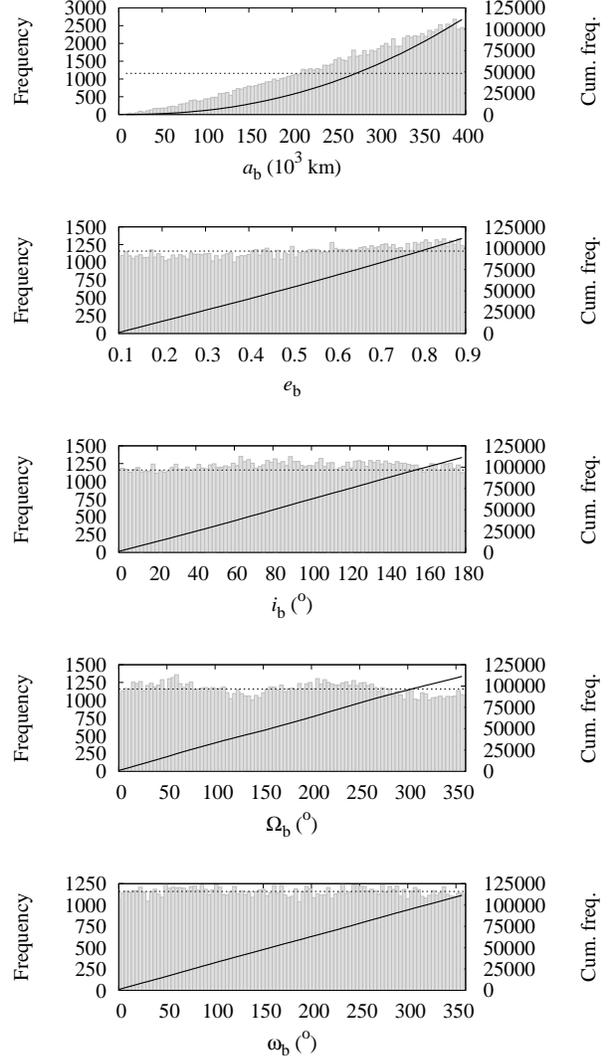}
        \caption{Frequency distribution of the initial orbital elements of the binaries that became disrupted couples after interacting 
                 with a trans-Plutonian planet (experiments as in Fig. \ref{BiDi}). The number of bins is 2 $n^{1/3}$ where $n=111\,307$. 
                 The dashed line shows the results of an equivalent uniform distribution. In this and subsequent histogram figures, the 
                 cumulative frequency is plotted as a black curve. 
                }
        \label{disini}
     \end{figure}
%
%-------------------------------------------------------------------------------------------------------------------------------------------
%
%
%-------------------------------------------------------------------------------------------------------------------------------------------
%
     \begin{figure}
        \centering
        \includegraphics[width=\linewidth]{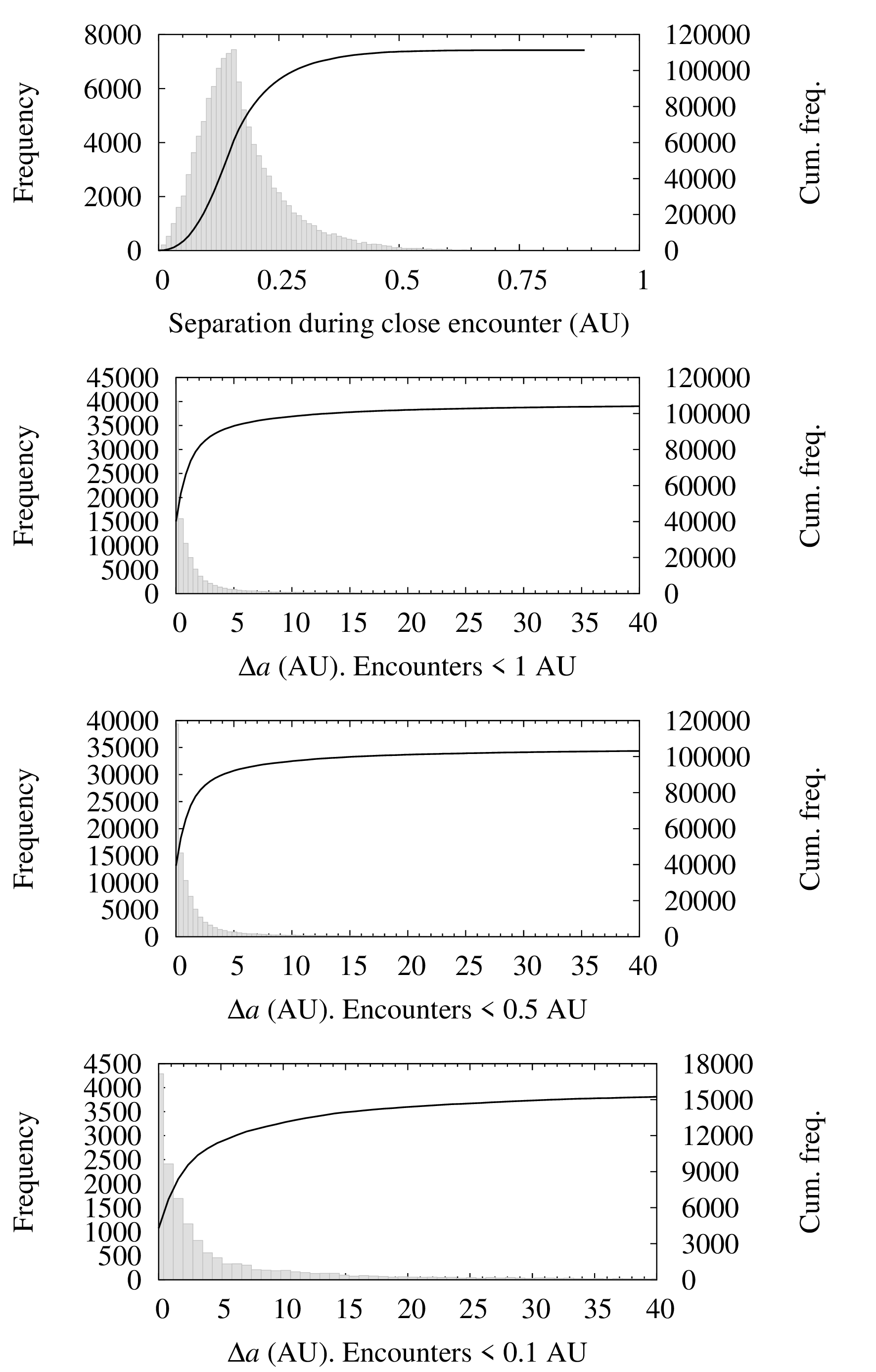}
        \caption{Frequency distribution of the separation during close encounter of the binaries analysed in Fig. \ref{disini} (top panel). 
                 Frequency distribution of the difference in heliocentric semi-major axes of the unbound couples 
                 (ejections/collisions/captures excluded) at the end of the simulation for encounters under 1 AU (second to top panel), 
                 0.5 AU (second to bottom panel), and 0.1 AU (bottom panel). The number of bins is 2 $n^{1/3}$ where $n$ is equal to 
                 111\,307, 106\,744, 105\,794 and 16\,489, respectively.  
                }
        \label{approdis}
     \end{figure}
%
%-------------------------------------------------------------------------------------------------------------------------------------------
%

\begin{figure}
        \centering
        \includegraphics[width=\linewidth]{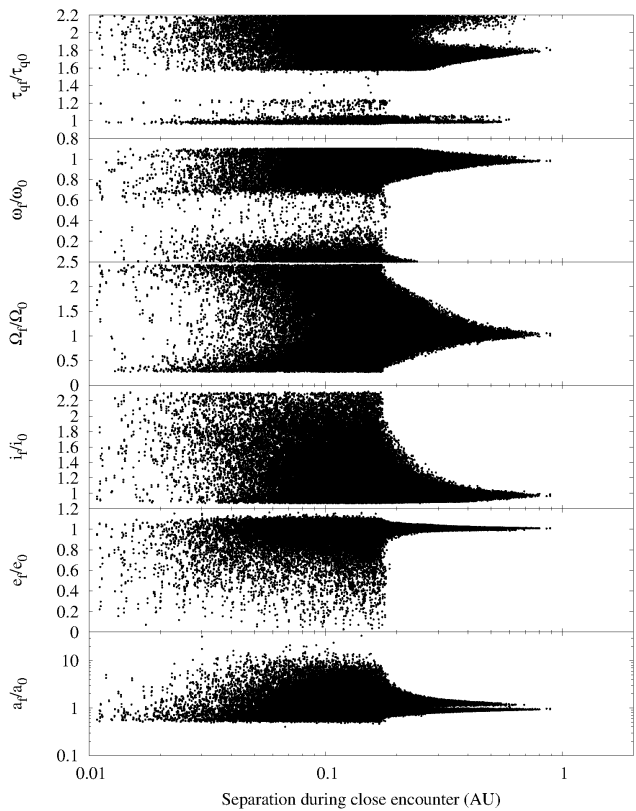}
        \caption{Ratio between the values of the orbital elements of the members of the unbound pair at the end of the simulation and the 
                 initial ones as a function of the separation during close encounter of the binaries (data as in Figs. \ref{disini} and 
                 \ref{approdis}, top panel, but excluding hyperbolic ejections and collisions/captures, 213\,488 unbound ETNOs). 
                }
        \label{closeR}
     \end{figure}

     Both Figs. \ref{BiDi} and \ref{BiDi++} show that newly disrupted couples may have relatively different values of the semi-major axis 
     and eccentricity but very similar values of the orbital inclination, longitude of the ascending node, and argument of perihelion 
     parameters. The difference in time of perihelion passage can range from weeks to centuries. This implies that unbound pairs are 
     expected to move initially along paths featuring similar directions of the perihelia, orbital poles, and perihelion/aphelion 
     velocities. Figure \ref{BiDi} shows the differences between the values of the orbital parameters, but in the case of the semi-major 
     axes, the actual values for many of the unbound pairs are very different from that of the parent binary. Figure \ref{closeR}, bottom 
     panel, shows that most disrupted pairs are scattered inwards. The orbital inclination tends to increase (Fig. \ref{closeR}, third panel 
     from bottom) and the value of the argument of perihelion decreases (Fig. \ref{closeR}, second panel). As for the statistical 
     significance of these results, let us consider as reference an isotropic distribution for the ratio of values (i.e. ratios $>1$ and 
     $<1$ are equally probable), where $\sigma=\sqrt{n}/2$ is the standard deviation of binomial statistics. The semi-major axes of the 
     unbound couples tend to be smaller than that of the parent binary; there is a 44$\sigma$ departure from an isotropic distribution in 
     $a_{\rm f}/a_{0}$. The resulting eccentricities tend to be smaller at the 9$\sigma$ level. However, the orbital inclinations of the
     members of the unbound pair do not show any statistically significant preference as they tend to increase at the 1$\sigma$ level. In 
     stark contrast, the longitude of the ascending node increases at the 90$\sigma$ level, the argument of perihelion decreases at the
     185$\sigma$ level, and the time of perihelion passage increases at the 164$\sigma$ level. Figure \ref{closeR} shows that closer 
     encounters tend to increase significantly the dispersion in the relative values of the orbital parameters; this trend seems to ease
     for very close encounters ($<0.1$~AU), but the number of points is too low to arrive at solid conclusions. 

\begin{figure}
        \centering
        \includegraphics[width=\linewidth]{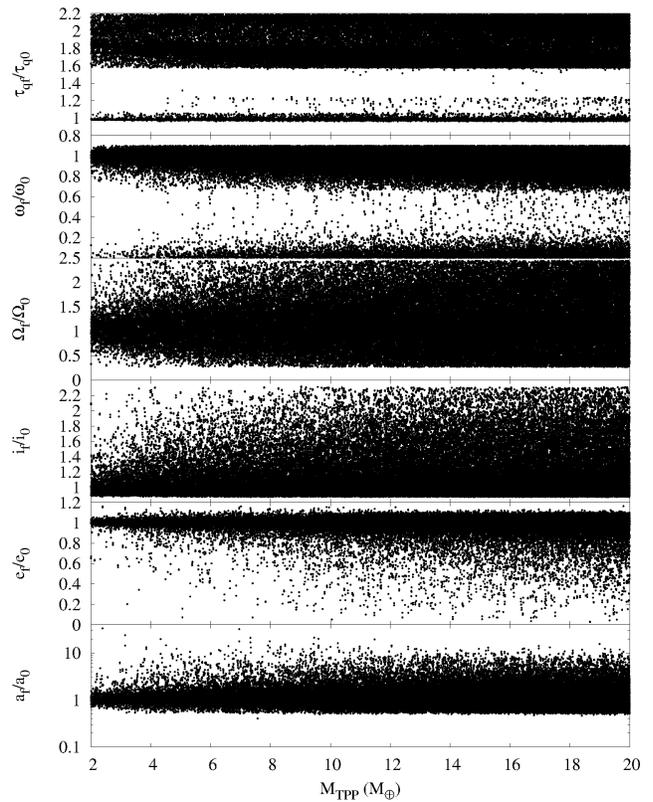}
        \caption{Same as Fig. \ref{closeR} but as a function of the mass of the trans-Plutonian planet.
                }
        \label{massR}
     \end{figure}

\begin{figure}
        \centering
        \includegraphics[width=\linewidth]{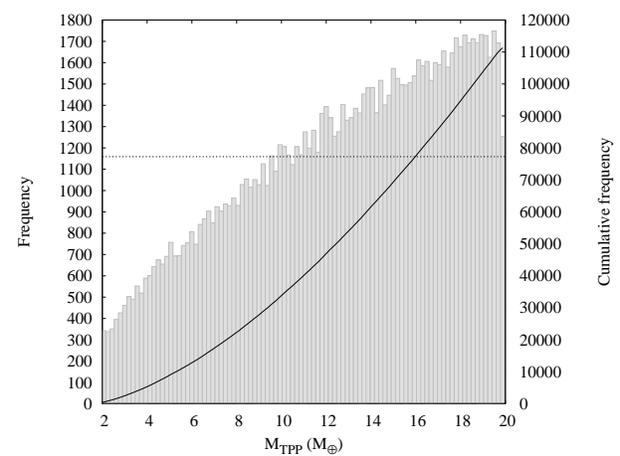}
        \caption{Frequency distribution of the mass of the perturber ($M_{\rm TPP}$) responsible for the binary dissociation events. The 
                 dashed line shows the results of a uniform distribution. Data as in Fig. \ref{approdis}, top panel.
                }
        \label{massdis}
     \end{figure}

     Figure \ref{massR} shows the ratio between the values of the orbital elements of the members of the unbound pair at the end of the 
     simulation and the initial ones as a function of the mass of the perturbing trans-Plutonian planet. The dispersion increases with the 
     mass of the perturber. The effect of the mass of the perturber is unclear from Fig. \ref{BiDi}, but it is important as we can see in 
     Fig. \ref{massdis}. The dashed line shows the results of the effectiveness of the binary dissociation process when the effect of the 
     mass of the perturber is negligible, in sharp contrast the actual distribution is far from uniform. Heavier perturbers are more 
     effective at disrupting binaries, all the other parameters being nearly the same. The mass of the putative perturber that triggered the 
     dissociation of the original binary may have been $>10$~$M_{\oplus}$ (see Fig. \ref{massdis}).
%
%-------------------------------------------------------------------------------------------------------------------------------------------
%
%
%-------------------------------------------------------------------------------------------------------------------------------------------
%
%
%-------------------------------------------------------------------------------------------------------------------------------------------
%
%
%-------------------------------------------------------------------------------------------------------------------------------------------
%
%
%-------------------------------------------------------------------------------------------------------------------------------------------
%
%
%-------------------------------------------------------------------------------------------------------------------------------------------
%

     Regarding the similarity in absolute terms between the parameters of the unbound pairs and those of the pair of ETNOs 
     474640--2013~RF$_{98}$, only a few unbound couples have values of their semi-major axes similar ($\pm$50~AU) to those of the ETNO pair 
     of interest here (see Table \ref{elements}). This suggests that, in this case, the hypothetical parent binary may have followed an 
     orbit somewhat exterior to those of the current pair of ETNOs. The orbit would have been more eccentric and perhaps slightly less 
     inclined.  

  \section{From the newly disrupted state to ETNO pair}
\begin{figure}
        \centering
        \includegraphics[width=\linewidth]{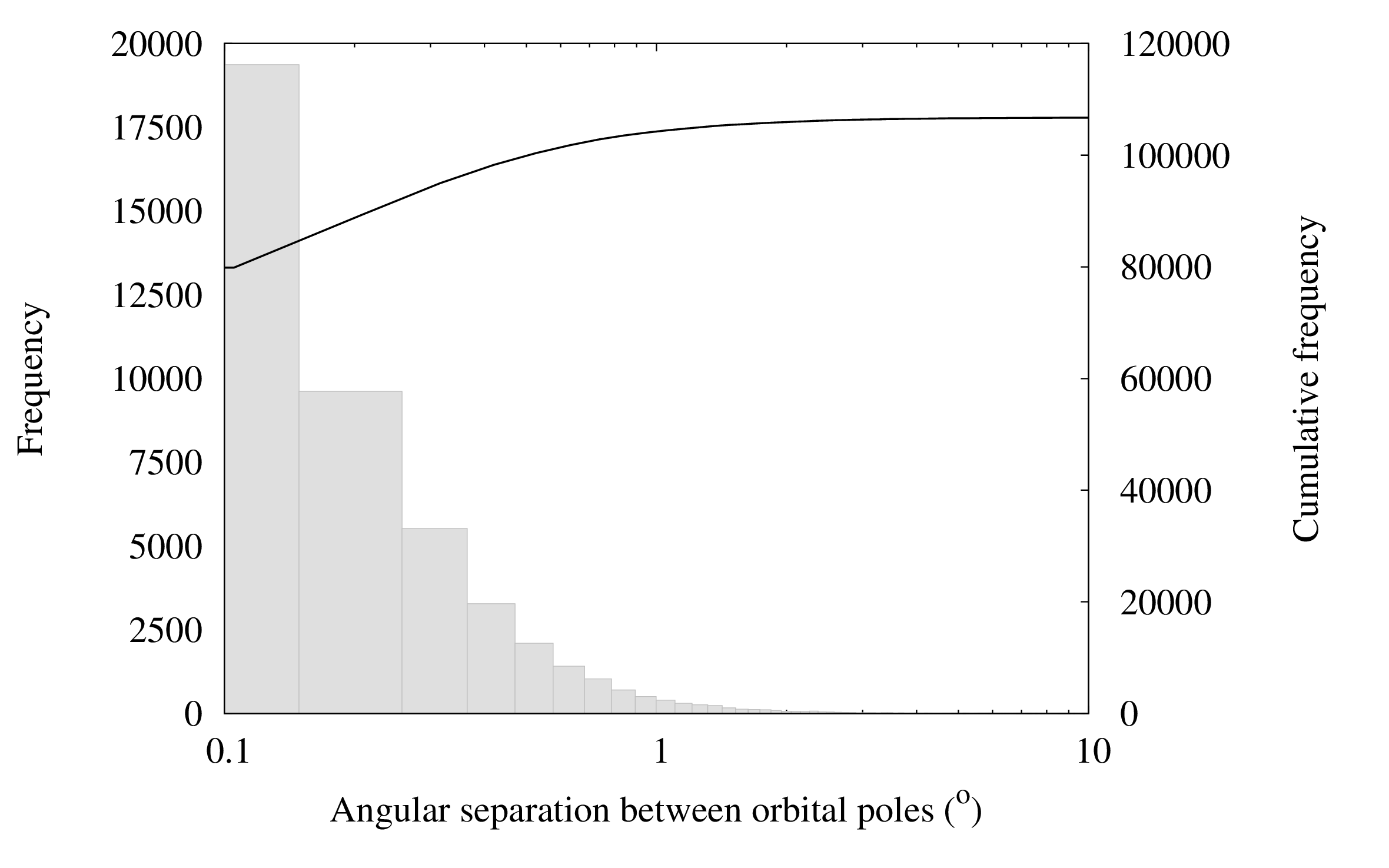}
        \includegraphics[width=\linewidth]{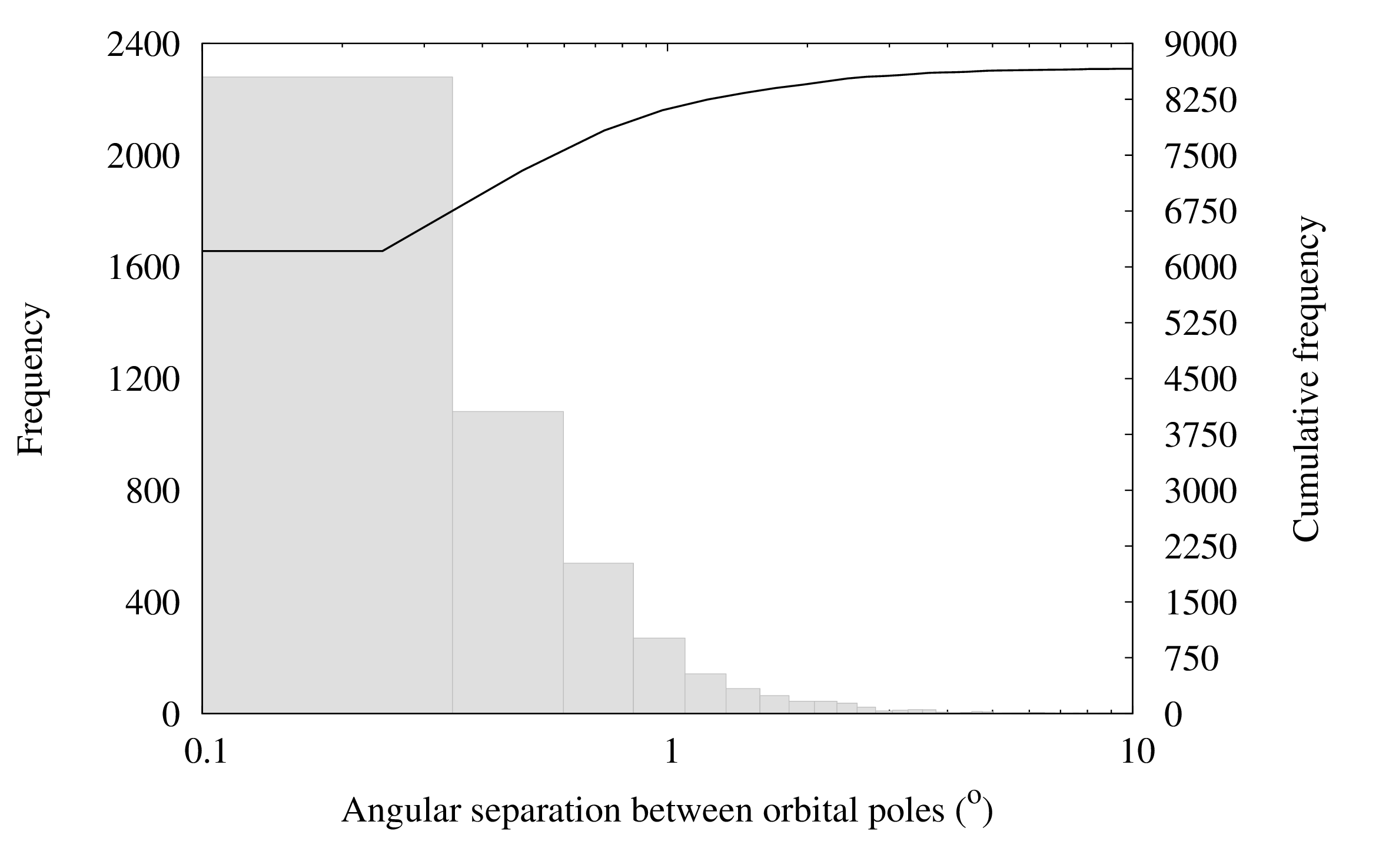}
        \caption{Frequency distribution of the angular separation between the orbital poles of the unbound pairs at the end of the 
                 simulation: all pairs (top panel) and pairs with $\Delta{a}>10$~AU (bottom panel). Data as in Fig. \ref{closeR}.
                }
        \label{angdis}
     \end{figure}

     The results obtained in the previous section show that newly disrupted couples resulting from binary dissociation events may have 
     relatively different values of $a$ and $e$ but very similar values of $i$, $\Omega$ and $\omega$; therefore, the unbound pairs move 
     initially along paths featuring similar directions of the orbital poles. Figure \ref{angdis}, top panel, shows the frequency 
     distribution of the angular separation between the orbital poles of the unbound pairs at the end of the simulation; focusing on the 
     8\,670 unbound pairs with $\Delta{a}>10$~AU (Fig. \ref{angdis}, bottom panel) does not change the frequency distribution too 
     significantly, but there is indeed a trend to have larger polar separations. 

     The vast majority of newly disrupted couples start their dynamical lives with their orbital planes mutually tilted by an angle 
     $<1$\degr. Finding newly disrupted couples with their orbital planes mutually tilted by a wider angle is certainly not impossible 
     (Fig. \ref{angdis}, bottom panel), but it is indeed far less probable. Although the angular separations between the orbital poles of 
     known ETNO pairs that might have been former binaries are small, none of them are below 1{\degr} (see sect.~2 and fig.~2 in de la 
     Fuente Marcos and de la Fuente Marcos 2016c); therefore and in order to explain the observational values, we must assume that the 
     initially very small separation increases over time due to some external force. The natural choice for the source of such secular 
     perturbation is the planet responsible for the binary dissociation event if the unbound pair experiences additional, more distant 
     encounters with it over an extended period of time. In this scenario, recurrent close approaches take place at regular intervals so we 
     can speak of resonant returns (Milani et al. 1999). Unfortunately, the perturbations from the very planet that caused the disruption 
     of the binary also make the determination of the past dynamical history of present-day unbound couples quite difficult.
%
%-------------------------------------------------------------------------------------------------------------------------------------------
%
%
%-------------------------------------------------------------------------------------------------------------------------------------------
%

\begin{figure}
        \centering
        \includegraphics[width=\linewidth]{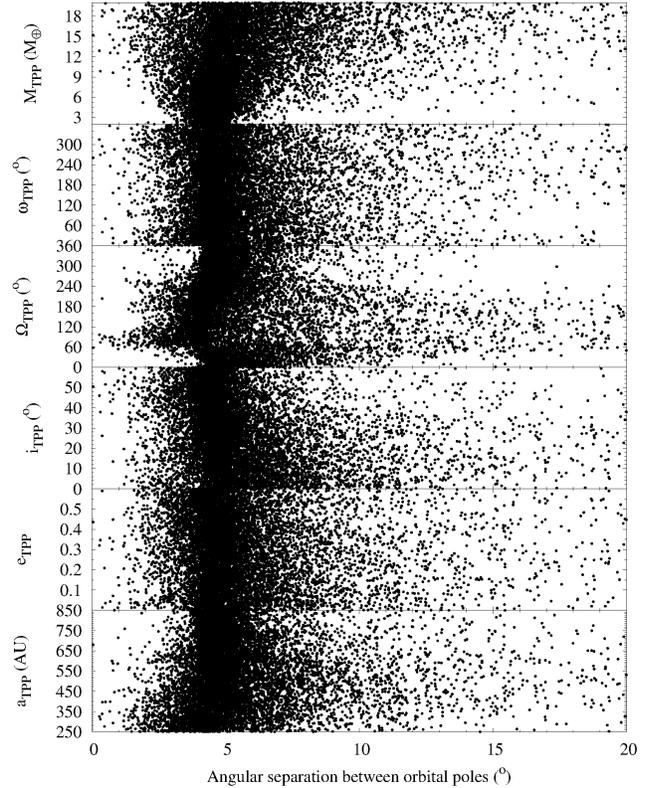}
        \caption{Orbital elements and mass of a sample of trans-Plutonian planets (TPPs) undergoing close encounters with the pair 
                 (474640) 2004~VN$_{112}$--2013~RF$_{98}$. The x-axis shows the angular separation between the orbital poles of the pair at 
                 the end of a simulation backwards in time for 8 Myr. The results of 11\,000 experiments are plotted.
                }
        \label{anglesb}
     \end{figure}

     In order to explore further the feasibility of this hypothesis, the numerical experiments carried out in this section do not involve 
     binaries but the actual ETNOs whose orbits are integrated backwards in time to study the evolution of the value of the angular 
     separation between their orbital poles when the ETNOs are subjected to the action of a sample of perturbers. This approach aims at 
     finding the most probable orbital parameters of a hypothetical planet able to tilt the orbital plane of the pair (474640) 
     2004~VN$_{112}$--2013~RF$_{98}$ from an initial angular separation close to zero at dissociation (see Fig. \ref{angdis}) to the 
     current value of slightly over 4\degr. Such experiments involve $N$-body integrations backwards in time under the influence of an 
     unseen perturber with varying orbital and physical parameters (assuming uniform distributions) so the relevant volume of parameter 
     space is reasonably well sampled. For these experiments, the ranges of the parameters of the perturber are the ones in 
     Fig. \ref{anglesb} and coincide with those in Fig. \ref{close}; the input orbits of the ETNO pair are based on the orbital elements in 
     Table \ref{elements} and computed using the MCCM method (see above) as in the case of the binaries in Sect. 4. The software applied and 
     the physical model assumed are also the same.
%
%-------------------------------------------------------------------------------------------------------------------------------------------
%
%
%-------------------------------------------------------------------------------------------------------------------------------------------
%

\begin{figure}
        \centering
        \includegraphics[width=\linewidth]{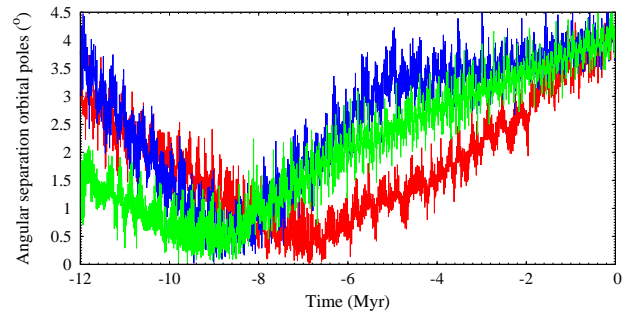}
        \caption{Three representative examples of the evolution of the angular separation between the orbital poles of the pair (474640) 
                 2004~VN$_{112}$--2013~RF$_{98}$ subjected to different perturbers. Red: $a$ = 348 AU, $e$ = 0.12, $i$ = 49\degr, 
                 $m$ = 20 $M_{\oplus}$. Blue: $a$ = 510 AU, $e$ = 0.17, $i$ = 34\degr, $m$ = 19 $M_{\oplus}$. Green: $a$ = 410 AU, 
                 $e$ = 0.09, $i$ = 49\degr, $m$ = 19 $M_{\oplus}$.
                }
        \label{examples}
     \end{figure}

     Figure \ref{anglesb} shows the results of 11\,000 eight-body experiments where the orbits of the pair of ETNOs were integrated 
     backwards in time for 8 Myr, subjected to the perturbation of a sample of trans-Plutonian planets. The impact of the properties of the 
     perturber on the final value of the angular separation between the orbital poles of the pair shows some interesting trends. Perturbers 
     with large values of the semi-major axis are somewhat disfavoured (see Fig. \ref{anglesb}, bottom panel); the best value could be 
     $\sim$350~AU. Eccentricities around 0.2--0.4 are preferred as well as orbital inclinations above 20{\degr} (see Fig. \ref{anglesb}, 
     third and second last panels). The longitude of the ascending node of the orbit of the perturber must be in the neighbourhood of 
     70{\degr} (see Fig. \ref{anglesb}, third panel) and the value of its argument of perihelion is left relatively unconstrained (see 
     Fig. \ref{anglesb}, second panel) although values in the neighbourhood of 60{\degr} and 280{\degr} seem to be preferred. The 
     mass of an effective putative perturber must be $>9$~$M_{\oplus}$ (see Fig. \ref{anglesb}, top panel) which is consistent with the 
     results in Sect. 4. Some representative examples of individual experiments are shown in Fig. \ref{examples} (see also fig. 5 in de 
     Le\'on et al. 2017). 

     The results of this set of experiments are only slightly different from those in Fig. \ref{close}; the largest differences appear in 
     the case of the values of eccentricity and argument of perihelion. However, being able to increase the tilt between two given orbital 
     planes and capable of triggering binary dissociation are not necessarily concurrent outcomes for a given set of initial conditions. 
     Although the process of disruption of wide ETNO binaries can only be effectively accomplished during binary-planet encounters at close 
     range (under about 0.25~AU, see above), smooth and progressive increase of the relative tilt does not require such level of proximity. 
     An unbound pair resulting from a close encounter may be inserted in new orbits that preclude further approaches at close range to the 
     planetary body that triggered the dissociation of the parent binary. In sharp contrast, our two sets of experiments use as input data 
     consistent orbits (with that of 474640) for both binaries and unbound pairs.

%
%-------------------------------------------------------------------------------------------------------------------------------------------
%
%
%-------------------------------------------------------------------------------------------------------------------------------------------
%

  \section{Discussion}
     The topic of the destruction of binaries by scattering encounters with a planet in the outer Solar System has been studied previously 
     by Parker and Kavelaars (2010). As ours, their work focuses on the end states of binary asteroids after binary-planet encounters, but 
     they simulate the effects of interactions between binary TNOs and Neptune. They used the \textsc{Mercury 6} $N$-body code (Chambers 
     1999) to integrate a population of 15\,000 particles for 1~Myr, performing 7\,500 integrations of the binary-TNO-Neptune interactions; 
     their binaries have binary semi-major axes uniformly distributed in the range 2\,000--120\,000~km, but the other binary orbital 
     parameters are similar to those in our experiments. In our longer integrations (24\,000 yr), nearly 95\% of the binary dissociations 
     correspond to unbound systems. Almost 99\% of the binary dissociations observed in their calculations were due to the binary semi-major 
     axis being enlarged to more than one Hill radius, not because the total energy of the system became greater than zero as in our case. 
     Such a sharp difference could be the result of using orbits of very different sizes; our ETNOs have $a\sim$328~AU, their TNOs have $a$ 
     in the range 20--34~AU. The fraction of collisions in both studies is about the same. Up to 80\% of binaries with binary semi-major 
     axis in units of the Hill radius $\sim$0.1 are disrupted in their simulations. In spite of the different scenarios implied, our 
     results are somewhat consistent with theirs; we find that binaries with $a_{\rm b}$ in units of the Hill radius $>0.14$ are 
     preferentially disrupted, although our disruption rates are lower than theirs probably because our integrations are much shorter. They 
     found that the probability of binary dissociation depends weakly ($<10$\%) on the initial eccentricity and inclination of the binary 
     pair which is consistent with our own findings, but we also find a weak dependency on the value of the mutual longitude of the 
     ascending node and argument of perihelion (see Fig. \ref{disini}).

\begin{figure}
        \centering
        \includegraphics[width=\linewidth]{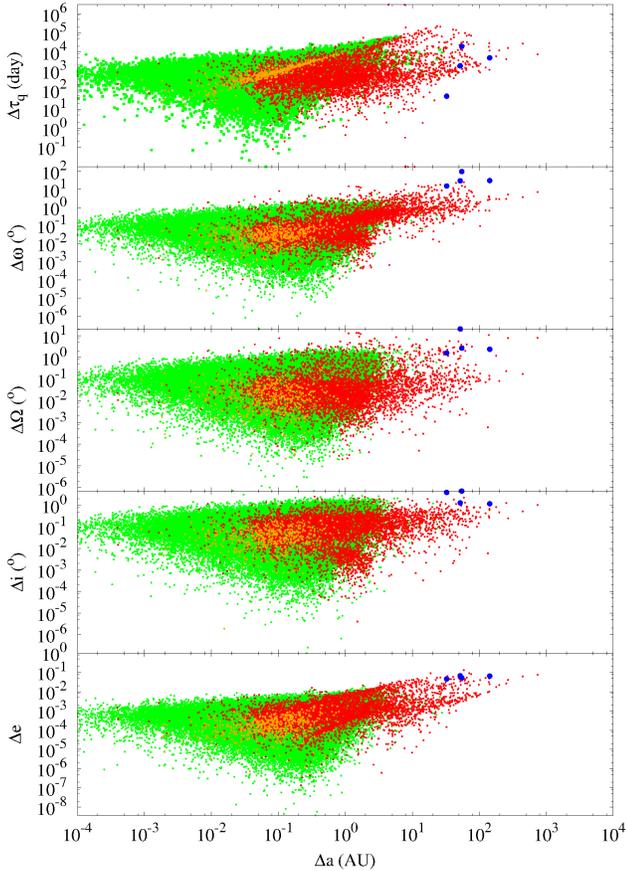}
        \caption{As Fig. \ref{BiDi} but for a different perturber (see the text for details). The results of 500\,000 experiments are 
                 plotted.
                }
        \label{HiIn}
     \end{figure}

     The results in Sect. 4 are based on a single representative orbit of the putative perturber; this orbit is only marginally compatible 
     with the orbital solution favoured in Sect. 5. But these results, how do they compare with those from a perturber moving in other 
     orbits? Figure \ref{HiIn} is similar to Fig. \ref{BiDi} but now the orbital parameters of the perturber are more consistent with the 
     results in Sect. 5 (see Fig. \ref{anglesb}). The orbit of the perturber ---$a=478.92\pm0.03$~AU, $e=0.338\pm0.003$, 
     $i=26.46\pm0.04$\degr, $\Omega=65.98\pm0.03$\degr, $\omega=273.853\pm0.013$\degr, and $f=148.41\pm0.06$\degr--- has been obtained after 
     performing a Monte Carlo-powered search analogous to the one producing the orbit of the perturber used in Sect. 4; its mass is assumed 
     to be in the range 2--20~$M_{\oplus}$. Each numerical experiment runs for 10\,500 yr or nearly one orbital period of the perturber. The 
     encounters now take place at about 474~AU from the Sun. For identical target binary population, this perturber is significantly less 
     efficient in triggering binary dissociations; only 1.9\% of the binaries were disrupted and out of them 91.2\% had positive relative 
     energy. The relative orbital elements of the resulting unbound pairs also exhibit distinctive features; in particular, unbound pairs 
     with differences in their times of perihelion passage below 1 yr are very scarce in the new simulations. 

     We also tested a more inclined orbit for the perturber ---$a=462.43\pm0.03$~AU, $e=0.1489\pm0.0009$, $i=51.08\pm0.02$\degr, 
     $\Omega=173.36\pm0.04$\degr, $\omega=78.26\pm0.03$\degr, and $f=75.24\pm0.04$\degr--- with each numerical experiment running for 
     10\,000 yr or nearly one orbital period of the perturber and encounters taking place at about 417~AU from the Sun. This particular
     perturber is precisely the one producing the smallest separation in Fig. \ref{close}. This set of calculations yielded a fraction of 
     destroyed binaries of nearly 0.9\%; over 88\% of them had positive relative energy. An additional set of experiments 
     ---$a=384.15\pm0.02$~AU, $e=0.528\pm0.002$, $i=27.31\pm0.03$\degr, $\Omega=100.52\pm0.03$\degr, $\omega=119.91\pm0.03$\degr, 
     and $f=67.27\pm0.02$\degr--- running for 8\,000 yr and producing encounters at about 209~AU from the Sun gave a binary disruption 
     rate of 9.2\% with 99.7\% of the disrupted binaries having positive relative energy at the end of the simulation. All these variations 
     are the result of the different geometry associated with the close encounters. The new perturbers require closer encounters to induce 
     effects comparable to those of the original one because the relative velocity during approaches at close range is now higher. In 
     general, fast binary-planet encounters are only disruptive if very deep; slower encounters can be effective even if they are 
     relatively shallow, but slow encounters are more sensitive to initial conditions.
%
%-------------------------------------------------------------------------------------------------------------------------------------------
%
%
%-------------------------------------------------------------------------------------------------------------------------------------------
%

     It may be argued that there are a host of other processes able to dissociate a binary system such as small impacts or solar tides; on 
     the other hand, dynamically-related pairs of asteroids can be the result of fragmentation at perihelion, strong mean motion or secular 
     resonances as well. Binary dissociation induced by asteroidal impacts requires a significant amount of debris to make it a viable 
     mechanism; it is unclear whether the outer Solar System between the trans-Neptunian belt and the Oort Cloud has the required amount of 
     mass orbiting at the right inclination. Solar tides are negligible unless the binary asteroid has a perihelion within a few tenths of 
     an astronomical unit (see e.g. Scheeres 2006). Fragmentation at perihelion is possible, but the absence of an obvious triggering 
     mechanism makes it unlikely.  

     The possible presence of former binaries among the known ETNOs has strong implications for the interpretation of the observed
     anisotropies in the distributions of the directions of their orbital poles and perihelia. Figures \ref{BiDi} and \ref{BiDi++} suggest 
     that a non-negligible fraction (perhaps higher than 25\%) of the known ETNOs might have had its origin in dissociated binaries. This 
     implies that their current orbital elements may correlate not just because of the secular perturbation of a putative trans-Plutonian 
     planet but as a result of a pre-existing dynamical link as well. This scenario also adds weight to the characterization of some of the 
     ETNOs as part of a transient population. A transitional nature, perhaps similar to that of the comets with $a<1\,000$~AU that are 
     interacting with Jupiter, appears to be consistent with the possible existence of a correlation between nodal distance and orbital 
     inclination in the case of both ETNOs and extreme Centaurs ($a>150$~AU but $q<30$~AU) as pointed out by de la Fuente Marcos and de la 
     Fuente Marcos (2017). In sharp contrast, many of the studies aimed at explaining the orbital architecture of the ETNO realm assume that 
     they are a long-term stable population. However, if they are a transient population, seeking dynamical mechanisms capable of making 
     them long-term dynamically stable may not be necessary. On the other hand, it is unclear whether the orbital diffussion scenario 
     recently proposed by Bannister et al. (2017) can produce pairs of orbitally correlated ETNOs, or not.

     Our results indicate that a planet with a mass in the range 10--20 $M_{\oplus}$ moving in a moderately eccentric (0.1--0.4) and 
     inclined (20--50\degr) orbit with semi-major axis of 300--600~AU, may be able to trigger the dissociation of a binary ETNO 
     following an orbit like the ones assumed here and induce a tilt similar to those observed on a time-scale of 5--10 Myr. Perturbers 
     with $M_{\rm TPP}<10\ M_{\oplus}$ or $a_{\rm TPP}>600$ AU are unable to produce the desired effects. Such a perturber should be 
     currently located well away from perihelion in order to have eluded detection by past surveys; this is to be expected in dynamical 
     terms as well, due to its eccentric orbit. On the other hand, the orbital solution that is most effective in triggering binary 
     stripping events actually reaches aphelion towards the Galactic plane, not far from the regions that surround the clouds of 
     Sagittarius, where the stellar density is the highest and outer Solar System surveys refuse to observe, to avoid a fog of false 
     positives. This probable coincidence reminds us that such perturber may be hidden in plain sight if it is currently moving projected
     towards those regions of the sky customarily avoided by surveys. Regarding its origin, planets similar to Uranus or Neptune 
     (super-Earths) may form at 125--750~AU from the Sun (Kenyon and Bromley 2015, 2016). Within this hypothetical context, smaller bodies 
     can also form in the same region prior to the actual planets ---perhaps even wide binaries. This scenario is however inconsistent with 
     the one proposed by Levison et al. (2008) that argues that most TNOs may have formed in the region interior to $\sim$35~AU and 
     subsequently scattered outwards by interactions with Neptune. Alternatively, such planet may have been scattered out of the region of 
     the Giant planets early in the history of the Solar System (Bromley and Kenyon 2016) or even captured from another planetary system (Li 
     and Adams 2016; Mustill et al. 2016) when the Sun was still a member of the open star cluster where it was likely formed.

     Although binary ETNOs have not yet been discovered, mechanisms capable of forming binaries in the outer Solar System have been 
     discussed in the literature (see e.g. Goldreich et al. 2002; Weidenschilling 2002; Astakhov et al. 2005; Schlichting and Sari 2008; 
     Nesvorn{\'y} et al. 2010). In fact and as pointed out above, Fraser et al. (2017a, 2017b) have found that the blue-coloured (spectral 
     slope $<17$\%), cold TNOs are predominantly in tenuously bound binaries and proposed that they were all born as binaries at 
     $\sim$38~AU. The pair of ETNOs discussed here are blue-coloured; the spectral slope of (474640) 2004~VN$_{112}$ is 12$\pm$2\% and that 
     of 2013~RF$_{98}$ is 15$\pm$2\% (de Le\'on et al. 2017). However, it is certainly too early to reach a final conclusion on the actual 
     place or places of origin of the ETNOs.

  \section{Conclusions}
     In this paper, we explore a dynamical pathway that may lead to the present-day pair of ETNOs (474640) 2004~VN$_{112}$--2013~RF$_{98}$ 
     and find that close encounters between extremely wide binary ETNOs and a trans-Plutonian planet can trigger binary dissociation. The 
     relative orbital properties of the unbound pairs resulting from these interactions resemble those of some documented ETNO pairs, 
     including the peculiar 474640--2013~RF$_{98}$. The presence of possible former binaries among the known ETNOs has profound 
     implications regarding the interpretation of the observed anisotropies in the distributions of the directions of the orbital poles and 
     perihelia of the ETNOs. Summarizing: 
     \begin{enumerate}[(i)]
        \item We confirm that the pair of ETNOs 474640--2013~RF$_{98}$ is an outlier in terms of relative orbital orientation within the 
              currently known sample of ETNOs.
        \item Wide binary ETNOs can dissociate during close encounters with putative trans-Plutonian planets. Most binary dissociation 
              events are the result of close encounters at minimum approach distances inside 0.25~AU. For encounters below 1~AU over 20\% 
              of the binaries become unbound pairs; nearly 75\% of all the binaries reaching separations from the perturber under 0.01~AU 
              are disrupted.
        \item Unbound pairs resulting from binary dissociation events may have relatively different values of the semi-major axis and 
              eccentricity but very similar values of the orbital inclination, longitude of the ascending node, and argument of perihelion
              parameters. The difference in time of perihelion passage ranges from weeks to centuries, but grows rapidly over time.
        \item The unbound pairs are expected to move initially along paths featuring similar directions of the perihelia, orbital poles, 
              and perihelion/aphelion velocities. 
        \item The unbound pairs can experience further, longer-range interactions with the perturber that may steadily increase their 
              relative inclination, longitude of the ascending node, and argument of perihelion. These changes will make the angular 
              separation between their orbital poles and perihelia progressively greater.
        \item The existence of former binaries among the ETNOs may signal the transient nature of many or all of them. If they are a 
              transient population, seeking dynamical mechanisms able to make them long-term dynamically stable may not be necessary.
     \end{enumerate}
     The research presented here must be understood as a proof-of-concept numerical exploration, not as an attempt at identifying the 
     actual parameters of the trans-Plutonian planet that probably triggered the formation of the unbound pair of ETNOs 
     474640--2013~RF$_{98}$. Multiple versions of the perturber (see Fig. \ref{close}) can lead to binary dissociation events similar to 
     the ones described here, but the actual probability of disruption depends strongly on the geometry of the encounter. Very precise 
     orbital solutions of the unbound pair under study are required to pursue a high-precision investigation in which the backwards 
     integration of their orbits subjected to the action of a perturber leads to a bound couple (assuming that they were originally bound). 

     Although the strength of disruptive encounters and the properties of the unbound couples depend on the choice of parameters and the 
     time interval, our full $N$-body investigation captures the essence of the binary dissociation mechanism and firmly establishes its 
     relevance within the ETNO context. As pointed out above, binary ETNOs have not yet been discovered and the known ETNOs have not been 
     thoroughly studied regarding binarity so they are assumed to be singles. Thus hypothetical wide, faint companions of ETNOs could be 
     challenging targets for the future. 

  \acknowledgments
     We thank the referee, J. A. Fern\'andez, for a critical, constructive and helpful review, an anonymous referee for another review, and 
     J. de Le\'on, J.-M. Petit, M. T. Bannister, D. P. Whitmire, G. Carraro, D. Fabrycky, A. V. Tutukov, S. Mashchenko, S. Deen and J. 
     Higley for comments on ETNOs and trans-Plutonian planets. This work was partially supported by the Spanish `Ministerio de Econom\'{\i}a 
     y Competitividad' (MINECO) under grant ESP2014-54243-R. CdlFM and RdlFM thank A. I. G\'omez de Castro, I. Lizasoain and L. Hern\'andez 
     Y\'a\~nez of the Universidad Complutense de Madrid (UCM) for providing access to computing facilities. Part of the calculations and the 
     data analysis were completed on the EOLO cluster of the UCM, and CdlFM and RdlFM thank S. Cano Als\'ua for his help during this stage. 
     EOLO, the HPC of Climate Change of the International Campus of Excellence of Moncloa, is funded by the MECD and MICINN. This is a 
     contribution to the CEI Moncloa. In preparation of this paper, we made use of the NASA Astrophysics Data System, the ASTRO-PH e-print 
     server and the MPC data server.

\end{document}